\documentclass[twocolumn]{autart}    
\usepackage{natbib}
\usepackage{har2nat}
\usepackage{amsmath}
\usepackage{amssymb}
\usepackage{bbm}
\usepackage{color}
\usepackage{epstopdf}
\usepackage{siunitx}
\usepackage{bm}
\usepackage{graphicx}
\usepackage{subfig}
\graphicspath{{figures/}}

\def\twon #1{\|#1\|}

\def\bR{\mathbb{R}}

\def\cC{\mathcal{C}}

\def\cI{\mathcal{I}}

\def\cL{\mathcal{L}}

\def\sgn{\text{sgn}}

\def\diag{\text{diag}}

\def \qed {\hfill \vrule height6pt width 6pt depth 0pt}
\def\bee{\begin{equation}}
	\def\ene{\end{equation}}
\def\beq{\begin{eqnarray}}
	\def\enq{\end{eqnarray}}
\def\bmatri{\begin{bmatrix}}
	\def\ematri{\end{bmatrix}}

\begin{document}
	\begin{frontmatter} 
		\title{The Isoline Tracking in Unknown Scalar Fields with Concentration Feedback} 
		\thanks[footnoteinfo] {This work was supported  in part by the National Natural Science Foundation of China under Grant 61722308.} 
		\author{Fei~Dong},
		\ead{dongf17@mails.tsinghua.edu.cn}
		\author{Keyou~You\corauthref{cor}}
		\corauth[cor]{Corresponding author} 
		\ead{youky@tsinghua.edu.cn}
		\address{Department of Automation and BNRist, Tsinghua University, Beijing 100084, China.}
		\begin{keyword}  Scalar field, isoline tracking, PI-like controller, regulation.
		\end{keyword} 
		
		\begin{abstract} 
			The isoline tracking of this work is concerned with the control design for a sensing vehicle to track a desired isoline of an unknown scalar field. To this end, we propose a simple PI-like controller for a Dubins vehicle in the GPS-denied environments. Our key idea lies in the  design of a novel sliding surface based error in the standard PI controller. For the circular field, we show that the P-like controller can {\em globally} regulate the vehicle to the desired isoline with the steady-state error that can be arbitrarily reduced by increasing the P gain, and is eliminated by the PI-like controller. For any smoothing field, the P-like controller is able to achieve the local regulation. 
			Then, it is extended to the cases of a single-integrator vehicle and a double-integrator vehicle, respectively. Finally, the effectiveness and advantages of our approaches are validated via simulations on the fixed-wing UAV and quadrotor simulators.   			
		\end{abstract}
	\end{frontmatter}
	
	\section{Introduction}
	The isoline tracking commonly refers to the tactic that a sensing vehicle reaches and then tracks a desired concentration level of a scalar field with unknown distribution, which has wide applications in the environmental exploration, e.g., tracking curve of sea temperature \citep{zhang2010cooperative}, tracking boundary of volcanic ash \citep{kim2017disturbance}, tracking plume front  of oil spill \citep{jiang2018plume}, exploring environmental feature of bathymetric depth \cite{mellucci2019environmental},  and monitoring algal bloom \citep{fonseca2019cooperative}. In the literature, it is also named as level set tracking \citep{Matveev2012Method}, curve tracking \citep{malisoff2017adaptive}, boundary tracking \citep{Menon2015Boundary,matveev2017tight,kim2017disturbance,mellucci2019environmental}, and covers the celebrated target circumnavigation as a special case \citep{Matveev2011Range,deghat2012target,Cao2015UAV,swartling2014collective,ZhengDistributed,dong2020Circumnavigating,lopez2020adaptive,dong2019Target}.
	
	Compared with the static sensor networks, it is more flexible and economical to utilize sensing vehicles to collect data or track targets. 		
	Roughly speaking, we can categorize the control methods for the isoline tracking depending on whether the gradient of the scalar field can be used or not. The gradient-based method is extensively used to steer a vehicle to track the direction of gradient descending (ascending) to the minimizer (maximizer) of a scalar field \citep{zhang2010cooperative,brinon2019multirobot,bourne2019coordinated}. This strategy can also be extended to the problem of the isoline tracking \citep{Kapitanyuk2018}.
	
	If the  gradient is not explicitly available, many works focus on the gradient estimation problem  \citep{brinon2019multirobot,hwang2019auv}, including that (a) one vehicle changes its position over time to collect the signal propagation at different locations; and (b) multiple vehicles collaborate to obtain measurements at different locations at the same time.  For example,  \citet{ai2016source} design a sequential least-squares field estimation algorithm for a REMUS AUV to seek the source of a hydrothermal plume. The stochastic method for extreme seeking is gradient-based in nature, the idea behind which is to approximate the gradient of the field by adding an excitatory input to the controller \citep{cochran20093,lin2017stochastic,LI2020Cooperative}. In \citet{brinon2019multirobot}, a circular formation of vehicles is adopted to estimate the gradient of the sensing field. Moreover, both cooperative Kalman filter and $\mathcal{H}_{\infty}$ filter are devised to estimate the gradient in  \citet{zhang2010cooperative} and  \citet{wu2012robust}, respectively. A particle filter has been developed to estimate  a Gaussian plume model where  multiple vehicles are coordinated via the multimodal nature of the nonparametric posterior in \citet{bourne2019coordinated}. 
	
	However, in many practical scenarios, the vehicles have no access to its GPS position and can only obtain the concentration measurement at the current location, i.e., the measurement is in a point-wise fashion \citep{Matveev2012Method}.  Thus, it is of interest to exploit gradient-free methods without position information. A sliding mode approach has been proposed for the target circumnavigation in \citet{Matveev2011Range} and then adopted to the level set tracking \citep{Matveev2012Method}, boundary tracking \cite{matveev2015robot}, and etc. They address the ``chattering" phenomenon via modeling dynamics of the actuator as a first-order linear differential equation. However, there is no rigorous proof of the revised control law. A PD feedback controller is devised in \citet{baronov2007reactive} for a double-integrator vehicle to follow isolines in a harmonic potential field. A PID controller with adaptive crossing angle correction is designed in \citet{newaz2018online}. Moreover, there are some heuristic methods for the isoline tracking, e.g., the sliding mode control \citep{Menon2015Boundary,mellucci2017experimental}, the bang-bang type control \citep{joshi2009experimental}, and etc. The sliding mode controller consists of two-sliding motions to explore the environmental feature of bathymetric depth. They validate their controller via simulations in a synthetic data-based environment and sea-trials via a C-Enduro ASV.  The bang-bang type control switches between alternative steering angles in virtue of whether the current measurement is above or below the threshold of interest, which results in a zigzagging behavior.
	
	In this paper, we propose a gradient-free controller in a PI-like form for a Dubins vehicle to track a desired isoline by using only the concentration feedback. That is, we do not use any field gradient or the position of the sensing vehicle, which is particularly useful in GPS-denied environments. Our key idea lies in the design of a novel sliding surface based error in the standard PI controller. Then we show that the steady-state tracking error can be reduced by simply increasing the P gain, and is eliminated for circular fields with a small I gain.  For the case of smoothing scalar fields, we explicitly show the upper bound of the steady-state tracking error, which also can be reduced by increasing the P gain. To validate the effectiveness of our PI-like controller via simulation, we adopt a fixed-wing UAV to track the predefined isoline of the concentration distribution of particulate matter (PM2.5) based on a real dataset in an area of China. Finally, we extend the PI-like controller to the cases of a single-integrator vehicle and a double-integrator vehicle, respectively.  A preliminary version of this work which only considers the case of a Dubins vehicle is presented in \citet{dong2020Coordinate}.

	The rest of this paper is organized as follows. In Section \ref{sec2}, we explicitly describe the isoline tracking problem. To solve it, we propose a simple PI-like controller in Section \ref{sec_controller}. In Section \ref{sec3}, we show the global convergence and local exponential stability for the case of circular fields.  In Section \ref{sec_saclar},  we study the closed-loop stability of the PI-like controller in a smoothing scalar field. The extension to the cases of a single-integrator vehicle and a double-integrator vehicle are given in Section \ref{sec_ext}. Finally, simulations are performed in Section \ref{secsim}, and some concluding remarks are drawn in Section \ref{sec6}.
	
	\section{Problem Formulation} \label{sec2}
	\begin{figure}[t!] 
		\centering{\includegraphics[width=0.9\linewidth]{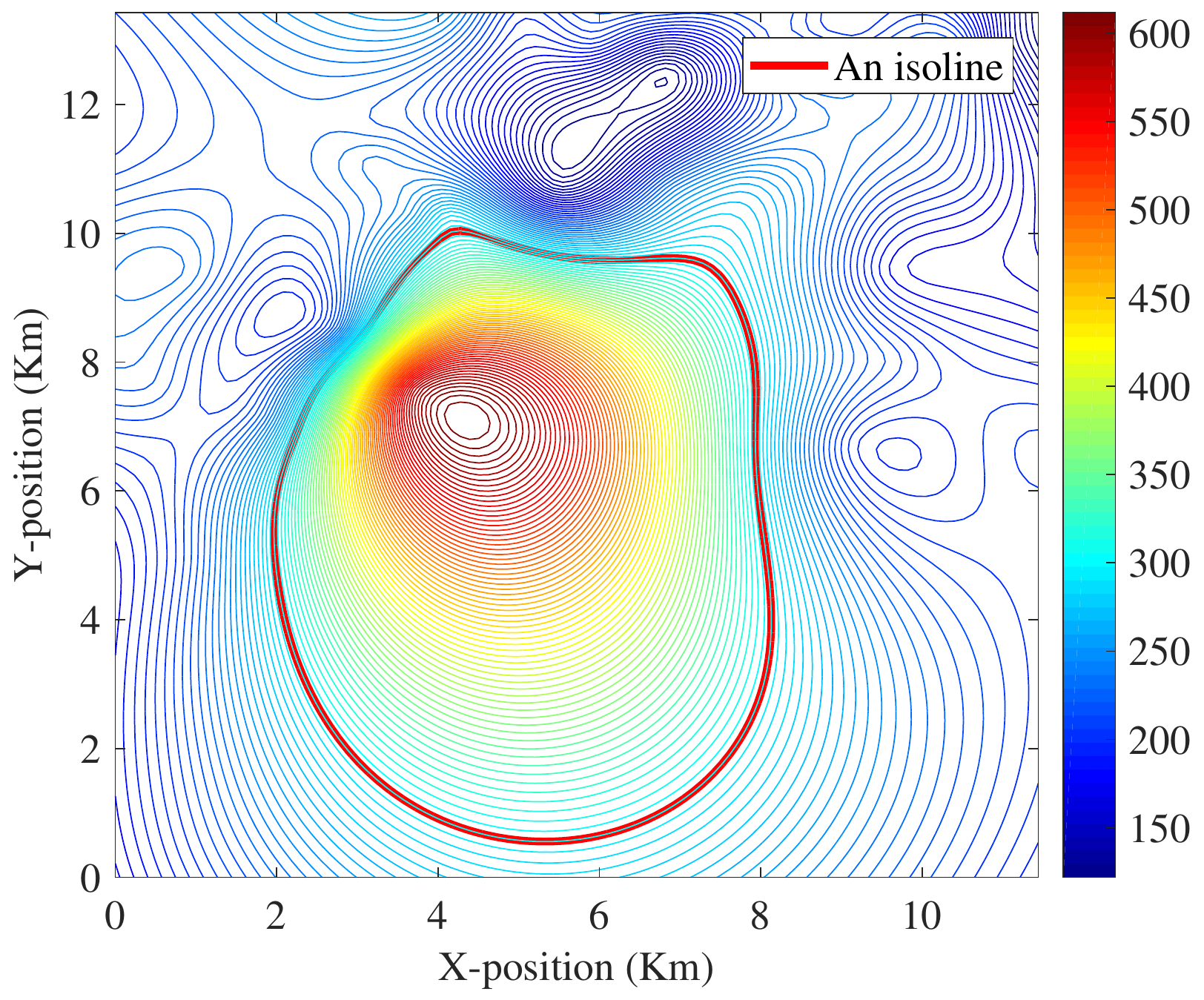}}
		\caption{The PM2.5 concentration observed in an area of China.}
		\label{fig1}
	\end{figure}
	In  Fig.~\ref{fig1}, we provide a 2-D example of the concentration distribution of PM2.5 based on a real dataset in an area of China\footnote{For privacy concern, we do not provide the exact region of the collected data.}. To monitor the environment, it is fundamentally important to investigate the spatial distribution of PM2.5. That is, we design a sensing vehicle to track an isoline of its distribution function, which is described as
	\begin{align} \label{eqsca}
		F(\bm p): ~\mathbb{R}^2 \rightarrow \mathbb{R},
	\end{align}
	where $\bm p \in \mathbb{R}^2$ is a GPS position in 2-D. 	
	Given a concentration level $s_d$,  an isoline $\cL(s_d)$ of $F(\bm p)$ is defined as  
	\begin{align} \label{eqset}
		\cL(s_d)=\{\bm p | F(\bm p) = s_d   \}.
	\end{align}
	
	The {\em isoline tracking} problem is on the control design for a sensing vehicle to move along with the desired isoline $\cL(s_d)$. Precisely, the position  $\bm p(t)$ of the sensing vehicle is controlled to satisfy that
	\begin{align} \label{eqobj}
		\lim\nolimits_{t\rightarrow \infty} |s(t)-s_d | \rightarrow 0~ \text{and} ~\|\dot{\bm p}(t)\| = v,
	\end{align}
	where $s(t)= F(\bm p(t))$ is the concentration of the scalar field  at the position $\bm p(t)$ and $v$ denotes a constant linear speed of the vehicle. Throughout this work, we always focus on the following scenario.
	\begin{enumerate}
	\renewcommand{\labelenumi}{\rm(\alph{enumi})}
\item Neither the concentration distribution function $F(\cdot)$ nor the GPS position of the vehicle $\bm p(t)$ is known. 
\item We cannot measure a continuum of the scalar field, and the vehicle can only obtain $s(t)$ at its current position $\bm p(t)$.
\item $s_d$ is not an extreme point of $F(\cdot)$.
\end{enumerate}

 The above implies that the gradient-based methods in \citet{zhang2010cooperative,malisoff2017adaptive,Kapitanyuk2018,brinon2019multirobot} cannot be applied. If $s_d$ is an extreme point of $F(\cdot)$, the isoline may be degenerated into a single point or a set with positive Lebesgue measure, in which case the isoline tracking problem is not well defined in this work. 
 
	\section{Controller Design for Dubins Vehicles} \label{sec_controller}
	In this section, we design a gradient-free controller in a PI-like form for a Dubins vehicle to complete the isoline tracking task. Our key idea lies in the design of a novel sliding surface based error in the standard PI controller.  The cases of a single-integrator vehicle and a double-integrator vehicle are given in Section \ref{sec_ext}.
	
	\begin{figure}[t!]
		\centering{\includegraphics[width=0.8\linewidth]{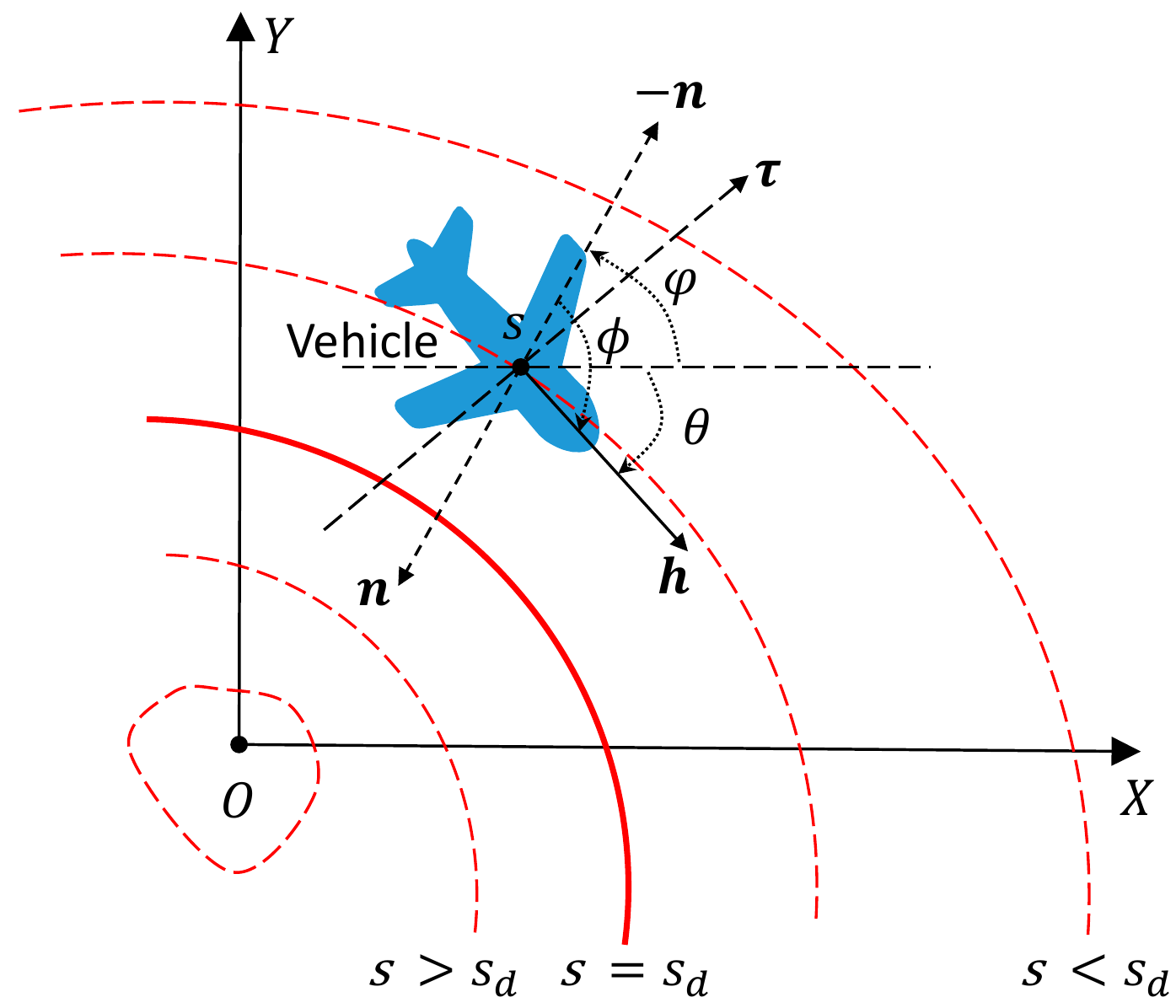}}
		\caption{Coordinates of the Dubins vehicle in scalar fields.}
		\label{fig_quad}
	\end{figure}
	Consider a Dubins vehicle on a 2-D plane
	\begin{equation} \label{eq1}
		\dot {\bm p}(t) = v \bmatri \cos \theta(t) \\ \sin \theta(t) \ematri \text{and} ~ \dot \theta(t) = \omega(t),
	\end{equation}
	where $\bm p(t) \in \bR^2$, $\theta(t)$, $\omega(t)$ and $v$ denote the GPS position, heading course, tunable angular speed and constant linear speed, respectively. See Fig.~\ref{fig_quad} for illustration.

	To achieve the tracking  objective in \eqref{eqobj} by the Dubins vehicle, we propose the following PI-like controller 
	\begin{align} \label{eq_Dubins}
		\omega(t)  = c_1 e(t) + c_2\sigma(t),
	\end{align}
	where $\sigma(t) = \int\nolimits_{0}^{t}e(\tau)\text{d}\tau$ is an integrator, $e(t)$ is the output of a nonlinear system driven by the tracking error $\varepsilon(t)=s(t)-s_d$, and $c_{1,2}\ge 0$ are the control parameters to be designed. 
	
	 The major difference of \eqref{eq_Dubins} from the PI controller lies in the novel design of the following error system
	\begin{align} \label{eqerror}
		e(t) = \dot \varepsilon(t) + c_3 \tanh ( {\varepsilon(t)}/{c_4}),
	\end{align}
	where $c_{3,4}>0$ are constant parameters and $\tanh(\cdot)$ is the standard hyperbolic tangent function.  
	In fact, $e(t)=0$ also can be regarded as a sliding surface. If the surface is maintained, i.e.,
	\begin{align}\label{eqsliding}
		\dot \varepsilon(t) =- c_3 \tanh \left(\varepsilon(t)/c_4 \right),
	\end{align}
	then $\varepsilon(t)$ will tend to zero with an exponential convergence speed.
	
	If we directly set $c_2=0$, then \eqref{eq_Dubins} is reduced to a P-like controller
		\begin{align} \label{eqp} 
		\omega(t)  = c_1 e(t).
		\end{align}
	Note that the P-like controller in  \eqref{eq_Dubins} is designed for the global stability, and the I-like controller is added to eliminate the steady-state error. Similar to the standard PI controller, it typically requires that $0\le c_2 \ll c_1$. In view of \eqref{eqsliding},  $c_3$ affects the convergence speed and $c_4$ affects the sensitivity to the tracking error $\varepsilon(t)$. 
	
	Since the PI-like controller \eqref{eq_Dubins} only uses the tracking error $\varepsilon(t)$ and its derivative $\dot \varepsilon(t)$, it is particularly useful in the GPS-denied environments.
	\begin{rem} If $\dot \varepsilon(t)$ is unavailable, we can design a second order sliding mode (SOSM) filter \citep{dong2020Circumnavigating}, a first order filter \citep{Guler2015Range}, or a washout filter \citep{Lin20163} to address this issue, which is not pursued in this work.  
	\end{rem}

	\section {The Isoline Tracking in Circular Fields} \label{sec3}
	In this section, we  consider a simplified yet instructive case of a circular field, which includes the acoustic field, i.e.,
	\bee F(\bm p)= I_0 \exp(-\alpha \Vert \bm p - \bm p_o\Vert _2)\label{circufield},\ene 
	where $\bm p_o$ is  the source position of the field and $I_0$, $\alpha$ are unknown positive parameters. Taking logarithmic functions on both sides of  \eqref{circufield}, then $\ln(F(\bm p)) = \ln I_o -\alpha \Vert \bm p(t) - \bm p_o \Vert _2$. Form the mathematical proof of view, there is no loss of generality to directly write the concentration function of a circular field as
		\begin{align}\label{eq_doubleield}
		 F(\bm p)=s_d- \alpha ( r(t) - r_d ),
	\end{align} 
where $r(t)=\Vert\bm p - \bm p_o\Vert _2$ is the distance from the vehicle to the source position, and $r_d$ is  unknown. Clearly, $F(\bm p)=s_d$ if and only if $r(t)=r_d$, and $r_d$ is desired distance for the vehicle to maintain from the source position $\bm p_o$.

By Fig. \ref{fig_quad}, let $\bm n = \nabla F(\bm p)$ denote the gradient of $F(\bm p)$ at the position $\bm p$, $\bm h =[\cos \theta, ~\sin \theta]'$ represent the heading vector of the vehicle, and $\bm \tau$ be a tangent vector of $\bm h$. By convention, $\bm h$ and $\bm \tau$ form a right-handed coordinate frame with $\bm h \times \bm \tau$ pointing to the reader.  Let $\phi(t)\in(-\pi, \pi]$ be the angle subtended by $-\bm n$ and $\bm h$, and $\varphi(t)\in(-\pi, \pi]$ is subtended by $-\bm n$ and the positive direction of $x$-axis. Without loss of generality, let the counter-clockwise direction of an angle be positive. Then, we have that $\phi(t) = \theta(t) - \varphi(t).$
	 
	 Now, we use $r(t)$ and  $\phi(t)$ to denote the coordinates of the polar frame centered at the source position $\bm p_o$. It follows from \eqref{eq_doubleield} that
	\begin{equation} \label{eqmodel}
		\begin{split}
			\dot s(t) &= -\alpha \dot r(t)=- \alpha v \cos \phi(t), \\
			\dot \phi(t) &= \omega(t) - \frac{v }{r(t)}\sin \phi(t) .
		\end{split}
	\end{equation}
	One can easily observe from Fig.~\ref{fig_quad} that $\omega(t)$ should be designed such that $[s_d,-\pi/2]'$ is a stable equilibrium of \eqref{eqmodel} to achieve the objective \eqref{eqobj}. At equilibrium it follows that
	\begin{align}
		\omega(t) = \omega_c = -v /r_d.
	\end{align}
	In the  circumnavigation problem,  $\omega_c$ is known for the controller design in \citet{dong2020Circumnavigating}. For the isoline tracking, this is not the case and we design an integrator $c_2 \sigma(t)$  in \eqref{eq_Dubins} to estimate $\omega_c$ which is indispensable for the exact isoline tracking.	
%
%
	
\subsection{The P-like controller}	
	Inserting \eqref{eqp} to \eqref{eqmodel} leads to that 
	\begin{equation} \label{eqmp}
	\hspace{-0.3cm}	\begin{split}
			\dot s(t) &= - \alpha v \cos \phi(t) ,\\
			\dot \phi(t) &= c_1 \left(\dot s(t) + c_3 \tanh \left( \frac{\varepsilon(t)}{c_4} \right) \right) - \frac{v \sin \phi(t)}{r(t)} .
		\end{split}
	\end{equation}
	
	One can show that \eqref{eqmp} has two equilibria, one of which is unstable  for any $c_{1,3,4}>0$ and of no interest. The other one is
	$\widetilde{\bm x}_e=[s_e, -\pi/2]'$ where   
	$s_e = s_d- \alpha (r_e -r_d)$
	and $r_e$ is the unique solution of $g(r)=0$ where 
		\begin{align} \label{eq_solu}
		g(r) := -\tanh \left( {{\alpha}(r -r_d)/{c_4}} \right) +{v}/{(c_1 c_3 r)}.
	\end{align}
Then, we have the following result.
\begin{lem}  \label{lemma_inc}
	Consider the equilibrium $\widetilde{\bm x}_e$, it holds that
	\begin{itemize}
		\item[(a)] $s_e<s_d$ for any finite $c_1>0$.
		\item[(b)]  $s_d-s_e$ decreases to zero as $c_1$ increases to infinity.
	\end{itemize}
	\end{lem}
	\begin{pf} By \eqref{eq_solu}, the proof is trivial.   \qed\end{pf}
	
Although the P-like controller in  \eqref{eqp}  is unable to {\em exactly} complete the isoline tracking task, the tracking error can be arbitrarily reduced by increasing the controller gain $c_1$, which is sufficient for application. We show below that the closed-loop system of \eqref{eqmp} converges {\em globally}  to $\widetilde{\bm x}_e $.
	
	\begin{prop} \label{prop_propo} 
	Consider the closed-loop system in (\ref{eqmp}) and let $\bm x(t)=[s(t), \phi(t)]'$.  If the controller parameters are selected to satisfy that
	\begin{align}  \label{eqcon}
	c_1 > 0, ~{\alpha} v>c_3>0, ~c_4>0,
	\end{align} 
	there exists a finite $t_1>t_0$ such that 
	\begin{align*}
	\twon{\bm x(t)-\widetilde{\bm x}_e}\le C \twon{\bm x(t_1)-\widetilde{\bm x}_e} \exp\left(-\rho  (t-t_1)\right), \forall t>t_1,
	\end{align*}
	where  $\rho$ and $C$ are two positive constants. 
\end{prop}
	\begin{pf}
		See Appendix.   \qed
	\end{pf}
	
	
		\begin{rem} If $c_3\ge v {\alpha}$, the Dubins vehicle either approaches the desired isoline $\cL(s_d)$ with oscillations or diverges from it. Since $\alpha$ is an unknown parameter, the vehicle can collect $N$ samples of $\dot s(t)$ and select $c_3$ such that
		\begin{align}\label{eqsample}
		c_3 < \frac{1}{N}\cdot \sum\nolimits_{i=1}^{N} |\dot s(i)|\le \alpha v,
		\end{align}
		where $i$ denotes the $i$-th sample.  
	\end{rem}
	By Lemma \ref{lemma_inc}, the steady-state error $s_d-s_e$ cannot be eliminated for a finite $c_1$.  This is where the integrator $\sigma(t)$ comes into play in the next subsection. 			
	
	
	%

\subsection{The PI-like controller for the exact isoline tracking}

	Inserting  \eqref{eq_Dubins} to \eqref{eqmodel}, we obtain that 
		\begin{equation} \label{eq24}
			\begin{split}
				\dot r(t) =&~  v \cos \phi(t), \\
				\dot \phi(t) =&  - c_1 \left(\alpha \dot r(t) +{c_3}\tanh \left({\alpha/{c_4}\cdot(r(t)-r_d)}\right) \right) \\
				&+c_2 \sigma(t)-{v \sin \phi(t)}/{r(t)} ,\\
				\dot \sigma(t)  =& -\alpha \dot r(t) - {c_3}\tanh \left({\alpha/{c_4} \cdot(r(t)-r_d)}\right).
			\end{split}
		\end{equation}
	 In \eqref{eq_Dubins},  the integrator $\sigma(t)$, which is sometimes called the internal model \citep[Chapter 12.3]{Khalil2002Nonlinear},  is designed to enforce the tracking error to converge to zero. Together with \eqref{eqsliding},  the objective in \eqref{eqobj} is finally achieved. Unfortunately, we only obtain the local regulation via linearization. As the P-like controller is already able to arbitrarily reduce this error, we do not pursue the nonlocal regulation result for the PI-like controller.


\begin{prop} \label{prop2}
	Consider the isoline tracking system in (\ref{eqmodel}) under the PI-like controller in \eqref{eq_Dubins}.  If the control parameters are selected to satisfy that
	\begin{align}  \label{eqcon2}
		c_1(c_1-2) v{\alpha}> c_2~\text{and} ~v{\alpha}>c_3>0,
	\end{align} 
	then $[r_d, -\pi/2,\omega_c/c_2]'$ is a locally exponentially stable equilibrium of \eqref{eq24}.
\end{prop}
	\begin{pf} Clearly, $\bm z_e=[ r_d,-\pi/2,\omega_c/c_2]'$ is an equilibrium of \eqref{eq24}. 
%
		Define an error vector
		\begin{align*}
			\bm z(t) &= \bmatri z_1(t),~z_2(t),~z_3(t)\ematri ' \\
			&= \bmatri r(t)-r_d, ~\phi(t)+\pi/2, ~\sigma(t)-\omega_c/c_2\ematri' ,
		\end{align*} 
		and linearize \eqref{eq24} around $\bm z_e $. It follows that
		\begin{align} \label{eqliner}
			\dot {\bm z} (t) = A \bm z(t) ,
		\end{align}	
		where the Jacobian matrix $A$ is given by
		\begin{align*}
			A= \bmatri & 0 &  v & 0 \\  &-{c_1 c_3\alpha}/{c_4 }  - {\omega_c}/{r_d}  &- {c_1 v \alpha} & c_2 \\ &-c_3\alpha/c_4 & -v\alpha & 0 \ematri.
		\end{align*}	
		
		Let $\mu_1 = {c_1 c_3\alpha}/{c_4} + {\omega_c}/{r_d}$, $\mu_2 =c_1 \alpha (c_1 \alpha v \mu_1 - c_2 c_3 \alpha/ (2c_4))$, $\mu_3=\mu_1v/2 + c_2 \alpha v/2 $, and $\mu_4 =c_1 c_2 c_4 v \mu_1/c_3 - c_2^2/2 $.   Consider the following Lyapunov function candidate  
		\begin{align}\label{eqvz}
			V(\bm z) = \bm z' P \bm z,
		\end{align}
		where $P$ is symmetric and obtained by
		\begin{align*}
			P =\frac{1}{2} \bmatri 2\mu_2 + \mu_1^2 & c_1 \alpha v \mu_1& -c_2 \mu_1 \\  c_1 \alpha v \mu_1 & 2 \mu_3 +(c_1 \alpha v)^2 & -c_1 c_2 \alpha v \\  -c_2 \mu_1 & -c_1 c_2 \alpha v & 2 \mu_4+ c_2^2   \ematri.
		\end{align*} 	
		
		One can verify  that the conditions in (\ref{eqcon2}) ensure the positiveness of $V(\bm z)$.  		
%
	Moreover,
		\begin{align} \label{eqvp}
			  V(\bm z) \le \lambda_M(P) \Vert \bm z \Vert_2^2,
		\end{align}	
		where $\lambda_{M}$ denotes the maximum eigenvalue of $P$.
		
		Then, taking the derivative of $V(\bm z)$ in \eqref{eqvz} along with \eqref{eqliner} leads to that 
		\begin{align}\label{eqq}
			\dot V(\bm z) = -\bm z' Q \bm z  ~\text{and}~	Q = \bmatri q_{11} & 0 & 0\\ 0 & q_{22} & q_{23} \\ 0 & q_{32} & q_{33}  \ematri,
		\end{align}
		where $q_{11} = c_1 \alpha v \mu_1^2 -{c_2 c_3 \alpha \mu_1}/{c_4} $, $q_{22} = (c_1 \alpha v)^3 $, $q_{23}=c_2 (c_1 \alpha v)^2 - c_2^2 \alpha v /2 + {c_1 c_2 c_4 (\alpha v)^2 \mu_1}/{(c_3 \alpha)}$, $q_{32}=q_{23}$, and $q_{33}= c_1 c_2^2 \alpha v$. Clearly, $Q$ is  positive definite. 
		It follows from \eqref{eqvp} and \eqref{eqq} that
		\setcounter{equation}{23}
		\begin{equation}
		\dot V(\bm z) \le- \lambda_m \Vert \bm z \Vert_2^2 \le-  {\lambda^{-1}_M}{ \lambda_m} V(\bm z),
		\end{equation}	
		where $\lambda_{m}$ denotes the minimum eigenvalue of $Q$.
		By the comparison principle \citep[Lemma 3.4]{Khalil2002Nonlinear}, $\bm z_e$ is a locally exponentially stable equilibrium of \eqref{eq24}.  \qed
	\end{pf}

	\section{The Isoline Tracking in Smoothing Fields}  \label{sec_saclar}
	In this section, we extend the circular field to more general cases satisfying the following assumption. 
	
	\begin{assum}\label{assum} The distribution function $F(\bm p)$ is twice continuously differentiable, and for any compact set $\Omega\subseteq \bR^2$ that excludes the stationary point of $F(\bm p)$, there exist  $\gamma_{1,2,3}>0$ such that   
	\begin{equation} \label{eqassum}
			\gamma_1\le \Vert \nabla F(\bm p) \Vert \le \gamma_2, ~\Vert \nabla^2 F(\bm p) \Vert \le \gamma_3,~ \forall \bm p\in\Omega.
	\end{equation}
	\end{assum}
	\begin{rem} Take the field in \eqref{circufield} as an example. Then, 
	\begin{align*}
		\Vert \nabla F(\bm p)\Vert = \alpha F(\bm p) ~\text{and}~
		\Vert \nabla^2 F(\bm p) \Vert = \alpha^2 F(\bm p),
	\end{align*} 
	when $\bm p \neq \bm p_o$. Obviously, (\ref{eqassum}) is satisfied.  
	\end{rem}
	Since $s_d$ is not an extreme point of $F(\bm p)$, it follows from Assumption \ref{assum} that the isoline $\cL(s_d)$ is composed by multiple strictly separate closed curve, i.e., $$\cL(s_d)=\bigcup\nolimits_{i\in\cI}\cC_i$$ where $\cI$ is a countable set, the set $\cC_i$ is a closed curve and $\cC_i\cap\cC_j=\emptyset$ if $i\neq j$. If $F(\cdot)$ is further convex, $\cL(s_d)$ contains only one closed curve. Otherwise, it may contain multiple disjoint closed curves, and the vehicle is expected to move along one of them, depending on the initial conditions.

	In view of Fig.~\ref{fig_quad}, we obtain that 
	\begin{align} \label{eq_dots}
			\dot s(t) = -v \Vert \nabla F(\bm p) \Vert \cos \phi(t).
	\end{align}
	Taking the derivative of $\dot s(t)$ leads to that
	\begin{align} \label{eq_ddots}
		\ddot s(t) 
		&= \omega(t) v \bm n' \bm \tau  +  v^2 \bm h '\nabla^2 F(\bm p) \bm h\\
		&= \omega(t) v \Vert \nabla F(\bm p) \Vert \sin \phi(t) +  v^2 \bm h '\nabla^2 F(\bm p) \bm h\nonumber.
	\end{align}
	By Fig.~\ref{fig_quad}, $\bm x_e= [s_d, -\pi/2]'$ is also the desired equilibrium of \eqref{eq_dots}.
	Suppose that $\bm x(t) = \bm x_e$, it follows from \eqref{eq_ddots} that
	\begin{align*}
		\ddot s(t) = -\omega(t) v \Vert \nabla F(\bm p) \Vert +  v^2 \bm h '\nabla^2 F(\bm p) \bm h.
	\end{align*}
	To maintain $\ddot s(t) =0$, it requires that 
	\begin{align} \label{eq_time}
		\omega(t)= \frac{ v \bm h '\nabla^2 F(\bm p) \bm h}{\Vert \nabla F(\bm p) \Vert},
	\end{align}
	which is time-varying and different from the case of the circular field of \eqref{circufield}. Since $F(\cdot)$ is unknown, we cannot use \eqref{eq_time}, which renders it impossible to exactly complete the isoline tracking task. Instead, we are able to design the P-like controller in \eqref{eqp} such that $|\varepsilon(t)|$ is uniformly bounded, and the bound can be arbitrarily reduced by increasing the P gain $c_1$. 
	\begin{prop} \label{prop_quad} Consider the isoline tracking system in \eqref{eq_dots} and \eqref{eq_ddots} under the P-like controller in \eqref{eqp}. Suppose that Assumption \ref{assum} holds and there is a closed curve in $\cL(s_d)$ such that $\phi(t_0) \in [-\epsilon, -\pi +\epsilon]$ where $\epsilon \in (0,\pi/2)$. Let the control parameters be selected to satisfy that
	\begin{align*}
		c_1 > \max\left\{\frac{\gamma_3 v}{\gamma_1\sin \epsilon \left ( v \gamma_1 \cos \epsilon - c_3  \right)}, ~ \frac{c_4\gamma_3 v +{c_3\gamma_2} }{  c_3\gamma_1  \sin \epsilon}\right\}, 
	\end{align*}
	and $0<c_3<v \gamma_1 \cos \epsilon$, then
	\begin{align*}
		\lim _{t\rightarrow \infty} |s(t) -s_d |\le \tanh^{-1} \left( \frac{c_4\gamma_3 v +{c_3 \gamma_2}  }{c_1c_3  \gamma_1  \sin \epsilon}  \right).
	\end{align*}
	\end{prop}
	The proof depends on the following technical result. 
	\begin{lem} \label{lemma_bound}
	Consider the following system
	\begin{align} \label{eq_bound}
		\dot z(t) = -k\tanh(z(t)) + b.
	\end{align}
	If $k>b>0$, then $\limsup\nolimits _{t \rightarrow \infty}|z(t) | \le \tanh^{-1}\left( {b}/{k}  \right).$
\end{lem}
	\begin{pf}
		Consider a Lyapunov function candidate as 
		\begin{align*}
			V_z(z) = {1}/{2}\cdot z^2(t).
		\end{align*}
		Taking the derivative of $V_z(z)$ along with \eqref{eq_bound} leads to that
		\begin{align*}
			\dot V_z(z) 
			& = z(t) \left (  -k\tanh(z(t)) + b \right)  \\
			&\le -k z(t) \tanh(z(t)) + b |z(t)|.
		\end{align*}	
		Since $k>b>0$, it holds that $\dot V_z(z)\le 0$ for all $|z(t)| \ge \tanh^{-1} \left({b}/{k} \right)$. This completes the proof.  \qed
		
	\end{pf}
	
	
	\begin{pf} [Proof of Proposition \ref{prop_quad}]
		Firstly, we  show that $\phi(t)$ cannot escape from the region $[-\epsilon, -\pi +\epsilon]$. To this end, inserting the P-like controller \eqref{eqp}   to \eqref{eq_ddots} leads to that
		\begin{equation} \label{eq_ddots1}
			\ddot s(t) =c_1 v \bm n' \bm \tau  ( \dot s(t) + c_3 \tanh ({\varepsilon(t)}/{c_4})  ) +  v^2 \bm h '\nabla^2 F(\bm p) \bm h .
		\end{equation}
		
		When $\phi(t) = -\epsilon $, it follows from \eqref{eq_ddots1} that
		\begin{align}\label{eqlower} 
			\ddot s(t) = &~ v^2 \bm h '\nabla^2 F(\bm p) \bm h  - c_1 v \Vert \nabla F(p) \Vert \sin \epsilon \times\nonumber \\
			&\left ( v \Vert \nabla F(p) \Vert \cos \epsilon + c_3 \tanh \left(  {\varepsilon(t)}/{c_4}\right)  \right) \nonumber  \\
			\le&- c_1 v \gamma_1\sin \epsilon \left ( v \gamma_1 \cos \epsilon - c_3  \right)  + \gamma_3 v^2\\
			<& 0. \nonumber
		\end{align}
		Similarly, $\phi(t) = -\pi +\epsilon$ leads to that 
		\begin{align} \label{equpper}
			\ddot s(t) 
			\ge &- c_1 v \gamma_1\sin \epsilon \left ( -v \gamma_1 \cos \epsilon + c_3  \right)  - \gamma_3 v^2  
			>  0.
		\end{align}	
		Since $\dot s(t)$ and $\phi(t)$ are continuous in $t$, then $\phi(t)$ will stay in the region $[-\epsilon,-\pi+\epsilon]$  if $\phi(t_0)\in[-\epsilon,-\pi+\epsilon]$. 
		
		Consider a Lyapunov function candidate as 
		\begin{align*}
			V_e(e) = {1}/{2}\cdot e^2(t).
		\end{align*}
		Taking the derivative of $V_e(e)$ along with \eqref{eq_dots} and \eqref{eq_ddots1} leads to that 
		\begin{align*}
			\dot V_e(e) 
			&= e(t) \left(\ddot s(t)  + {c_3}/{c_4}\cdot \left( 1- \tanh^2\left( { \varepsilon(t)}/{c_4} \right) \right) \dot s(t)   \right)\\
			&=    c_1 v \bm n' \bm \tau   e^2(t)  +e(t) \times\\
			& ~~~\left(   v^2 \bm h '\nabla^2 F(\bm p) \bm h  + {c_3}/{c_4}\cdot \left( 1- \tanh^2\left({ \varepsilon(t)}/{c_4} \right) \right) \dot s(t)   \right) \\
			&\le     c_1 v \bm n' \bm \tau   e^2(t)  + \left(\gamma_3 v^2 +{c_3}/{c_4}\cdot \gamma_2 v \right) |e(t)|        \\
			&\le    - \left(c_1 v \gamma_1  \sin \epsilon \right)  e^2(t)  + \left(\gamma_3 v^2 +{c_3}/{c_4}\cdot \gamma_2 v \right) |e(t)|.        
		\end{align*}
		It is clear that $\dot V_e(e)\le 0$ holds for all 
		\begin{align*}
			| e(t)| \ge \eta := \frac{\gamma_3 v +{c_3 \gamma_2}/{c_4}  }{c_1  \gamma_1  \sin \epsilon}.
		\end{align*}
		Thus, $|e(t)|$ will be eventually bounded by $\eta$, i.e., 
		\begin{align*}
			\lim\nolimits_{t\rightarrow \infty} \left| \dot s(t) + c_3\tanh \left ( {\varepsilon(t)}/{c_4}  \right) \right|  \le \eta.
		\end{align*}
		
		By Lemma \ref{lemma_bound} and the condition that $c_1>\frac{c_4\gamma_3 v +{c_3\gamma_2} }{ c_3\gamma_1  \sin \epsilon}$, it implies  
		\begin{align*}
			\lim _{t \rightarrow \infty}|s(t) - s_d| \le   \tanh^{-1} \left( \frac{c_4\gamma_3 v +{c_3 \gamma_2}  }{c_1c_3  \gamma_1  \sin \epsilon}  \right).
		\end{align*}	
		This completes the proof.    \qed
	\end{pf}

	\section{Extension to Other Vehicles} \label{sec_ext}
	In this section, we further extend the PI-like controller \eqref{eq_Dubins} for the Dubins vehicle \eqref{eq1} to the cases of a single-integrator vehicle and a double-integrator vehicle, respectively.
		
	\subsection{Controller design for single-integrator vehicles}
	Consider a single-integrator vehicle as follows
	\begin{equation} \label{eq_single}
		\begin{split}
			\dot {\bm p}_1(t) = {\bm v}_1(t),
		\end{split}
	\end{equation}
	where $\bm p_1(t)$ and $\bm v_1(t)$ denote the position and velocity of the single-integrator vehicle in 2-D, respectively. 
	
	To complete the isoline tracking task in \eqref{eqobj} by the single-integrator vehicle \eqref{eq_single}, we propose a {\em concentration-only} controller 
	\begin{align}\label{controller_sing}
		\bm v_1(t) = v \bmatri \cos \theta_1(t), &\sin \theta_1(t) \ematri',
	\end{align}
	where $v$ is the constant linear speed and $\theta_1(t)$ is given as
	\begin{align}\label{controller_single}
		\theta_1(t) = c_1 s(t) + c_1c_3 \zeta(t),~\dot \zeta(t) = \tanh\left({\varepsilon(t)}/{c_4}\right).
	\end{align}
	
	Taking the time derivative of $\theta_1(t)$ leads to that 
	\begin{align*}
		\dot \theta_1(t) =c_1\dot s(t) + c_1c_3 \tanh\left({\varepsilon(t)}/{c_4}\right)= c_1 e(t),
	\end{align*}
	which is of the same as the P-like controller \eqref{eq_Dubins}. In this case, the trajectories of the Dubins vehicle \eqref{eq1} and single-integrator vehicle \eqref{eq_single} are identical if they have same initial states, as shown in Lemma \ref{lemma_single}.
	
	\begin{lem} \label{lemma_single}
	Consider the Dubins vehicle \eqref{eq1}  under the P-like controller in \eqref{eqp} and the single-integrator vehicle \eqref{eq_single} under the controller \eqref{controller_single}. If the two vehicles start at the same initial states, i.e., $\bm p(t_0)=\bm p_1(t_0)$ and $\theta(t_0)=\theta_1(t_0)$, then their trajectories are  identical.
	\end{lem}
	\begin{pf}
		Define an error vector as follows
		\begin{align}\label{eqz1}
			\bm z(t) = [\bm p'(t)-\bm p'_1(t), & \theta(t)-\theta_1(t)]'.
		\end{align}
		If the Dubins vehicle \eqref{eq1} and the single-integrator vehicle \eqref{eq_single} have the same state at some time $t$, e.g., $\bm z(t)=\bm 0$, it further holds that
		\begin{align*}
			\dot{\bm z}(t) = [ \bm v'(t) -\bm v'_1(t), & \dot \theta(t)-\dot \theta_1(t) ]'= \bm 0,
		\end{align*}		
		where $\bm v(t)=v[\cos\theta(t), \sin\theta(t)]'$ is the velocity of the Dubins vehicle. Thus, the trajectories of the vehicles \eqref{eq1} and \eqref{eq_double} are identical if they have same initial states.    \qed
	\end{pf}
%

	\subsection{Controller design for double-integrator vehicles} \label{sec-pism}
	\begin{figure}[t!]
		\centering{\includegraphics[width=0.8\linewidth]{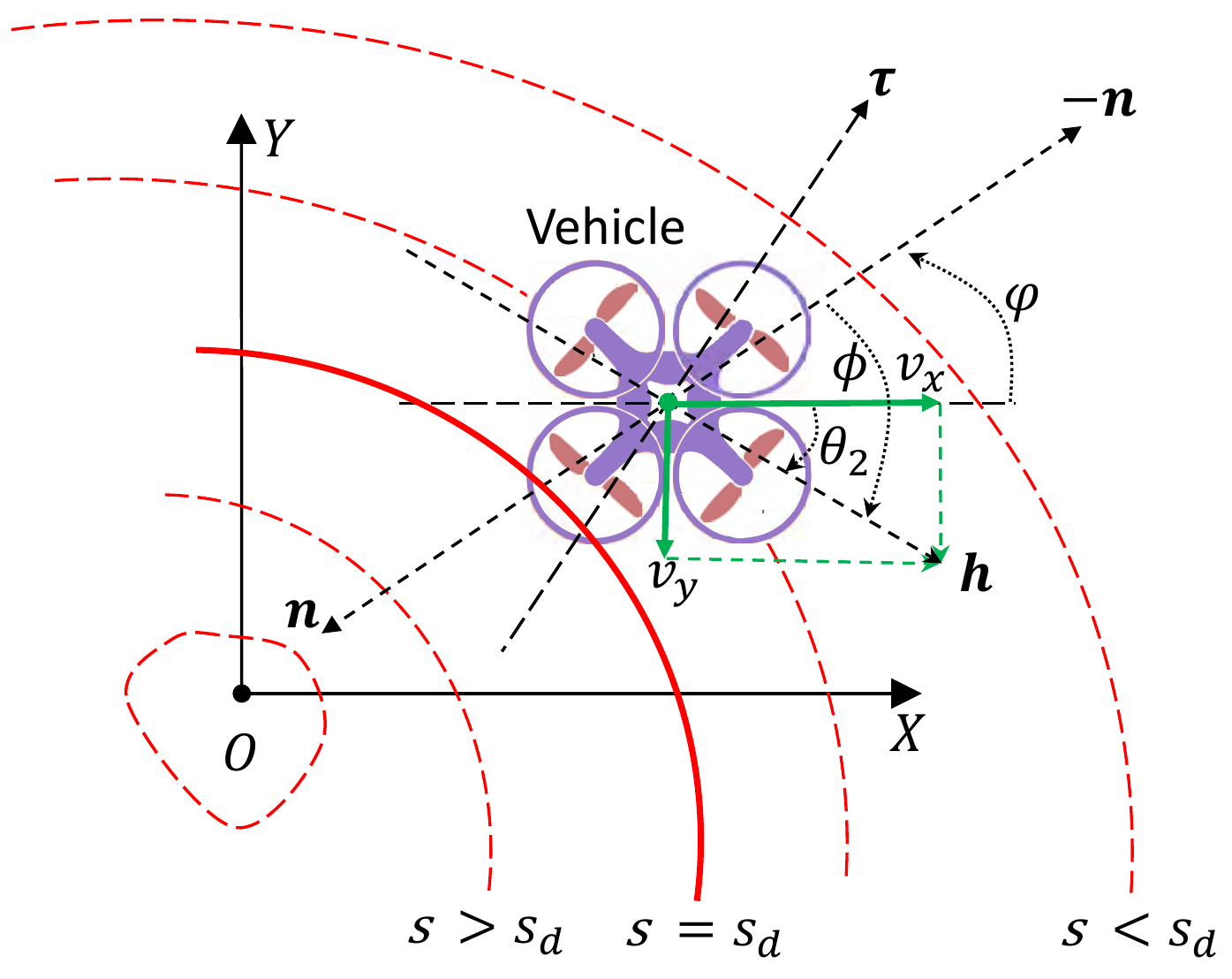}}
		\caption{Coordinates of the double-integrator vehicle in scalar fields.}
		\label{fig_quadrotor}
	\end{figure}
	Consider a double-integrator vehicle in Fig.~\ref{fig_quadrotor}
	\begin{equation} \label{eq_double}
		\begin{split}
			\dot {\bm p}_2(t) = \dot {\bm v}_2(t) ~\text{and}~\dot {\bm v}_2(t) = \dot {\bm a}_2(t),
		\end{split}
	\end{equation}
	where $\bm p_2(t)$, $\bm v_2(t)= [v_x(t), v_y(t)]'$, and $\bm a_2(t)$ denote the position, velocity, and acceleration in 2-D, respectively. 
	
	We propose the following controller
	\begin{align}\label{eq_Double}
		\bm a_2(t) = \underbrace{\omega(t)\bmatri- v_y(t)\\ v_x(t)\ematri}_{\text{For isoline tracking}}  +\underbrace{c_5\cdot \sgn\left(\bm v_2^d(t) - \bm v_2(t)\right)}_{\text{For velocity regulation}},
	\end{align}
	where  $\sgn(\bm x)$ returns the sign of each element of $\bm x$,  $c_5>0$ is the control parameter to be determined, $\omega(t)$ is the PI-like controller  in \eqref{eq_Dubins},  the  desired velocity is decomposed as
		\begin{align*}
		\bm v_2^d(t) = v [\cos\theta_2(t), &\sin\theta_2(t)]',
	\end{align*}
	and $\theta_2(t)=\arctan(v_y(t)/v_x(t))$ in Fig.~\ref{fig_quadrotor}.
	
 In  \eqref{eq_Double}, the first term is orthogonal to $\bm v_2(t)$ and is used to complete the isoline task and the other aims to regulate the velocity $\bm v_2(t)$ such that 
		\begin{align*}
			v_2(t) = v, ~\forall t>t_0+T,
		\end{align*}
		where $v_2(t)=\Vert \bm v_2(t) \Vert_2$ is the linear speed of the double-integrator vehicle, and $T>0$ is finite. 
	\begin{lem} \label{lemma_v2}
	Consider the double-integrator in \eqref{eq_double} under the controller \eqref{eq_Double}, there is a finite $T>0$ such that 
	\begin{align*}
		v_2(t) = v, ~\forall t>t_0+T.
	\end{align*}
	\end{lem}
	\begin{pf}
				Consider the following Lyapunov function   
		\begin{align*}
			V_v(v_2) = {1}/{2}\cdot (v_2(t)-v)^2. 
		\end{align*}
	Taking the derivative of $V_v(v_2)$ along with \eqref{eq_double} and \eqref{eq_Double} leads to that
		\begin{align*}
			\dot V_v (v_2) 
			=&{1}/{v_2(t)}\cdot (v_2(t)-v)\left( v_x(t)\dot v_x(t) + v_y(t)\dot v_y(t)  \right)\\
			=&-c_5|v_2(t)-v|  \left(|\cos \theta_2(t)|+|\sin \theta_2(t)|\right)\\
			\le&-c_5|v_2(t)-v|\\
			=&-\sqrt{2}c_5V_v^{1/2}(v_2).
		\end{align*}
		By the comparison principle, it follows that
		\begin{align*}
			v_2(t) = v, ~\forall t > t_0+ T,
		\end{align*}
		where $T =\sqrt{2} V_v^{1/2}(v_2(t_0))/c_5$.    \qed	
	\end{pf}
	
	After a finite time of length $T$, the double-integrator vehicle in \eqref{eq_double} is  only controlled by the first term of \eqref{eq_Double}
	\begin{align}\label{eq_Doublec}
		\bm a_2(t) = \omega(t) [ - v_y(t),  & v_x(t) ]'.
	\end{align}
	
	Since the above is orthogonal to $\bm v_2(t)$, we can show that the trajectories of the two vehicles \eqref{eq_Dubins} and \eqref{eq_double} are identical if they have same initial states.
	
	\begin{lem} \label{lemma_iden}
	Consider the Dubins vehicle \eqref{eq1} under the PI-like controller \eqref{eq_Dubins} and the double-integrator vehicle \eqref{eq_double} under the controller \eqref{eq_Double}. If the two vehicles have same initial states, i.e., $\bm p(t_0)=\bm p_2(t_0)$, $v_2(t_0)=v$, and $\theta(t_0)=\theta_2(t_0)$,  their trajectories are  identical.
	\end{lem}
	\begin{pf}
		Similar to \eqref{eqz1}, we define an error vector as  
		\begin{align*}
			\bm z(t) = [ \bm p'(t)-\bm p'_2(t), & \bm v'(t)-\bm v'_2(t)]',
		\end{align*}
		where $\bm v(t) = v[\cos \theta(t), \sin\theta(t)]'$.
		If the Dubins vehicle \eqref{eq1} and double-integrator vehicle \eqref{eq_double} have same states at some time $t$,  it holds that $\bm z(t)=\bm 0$ and $\dot {\bm z}(t) =\bm 0$.
		Thus, the trajectories of the vehicles \eqref{eq1} and \eqref{eq_double} are identical if they have same initial states.   \qed 
	\end{pf}
	
	
	\section{Simulations} \label{secsim}
	In this section, the effectiveness and advantages of the proposed controllers are validated by simulations. Particularly, the PI-like controller \eqref{eq_Dubins} and the controller \eqref{eq_double} are performed on the simulators of (a) a 6-DOF fixed-wing UAV in the field of PM2.5; and (b) a quadrotor built by CrazyFlie 2.0 platform, respectively.
	\subsection{The isoline tracking in a circular field} \label{sec_sub2}
\begin{figure}[t!]
		\centering{\includegraphics[width=0.8\linewidth]{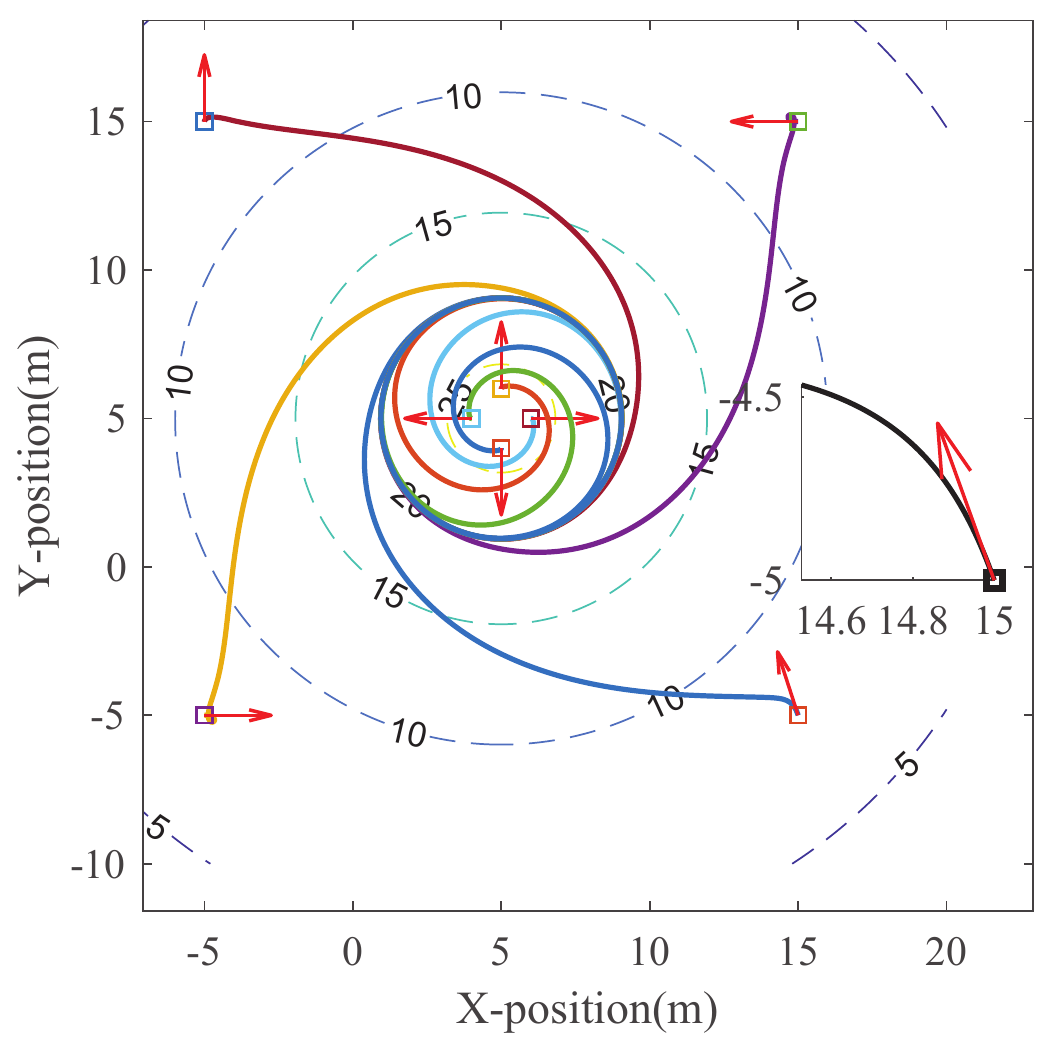}}
		\caption{Trajectories of the Dubins vehicle with different initial states.}
		\label{fig4}
	\end{figure}

\begin{figure}[t!]
		\centering{\includegraphics[width=0.8\linewidth]{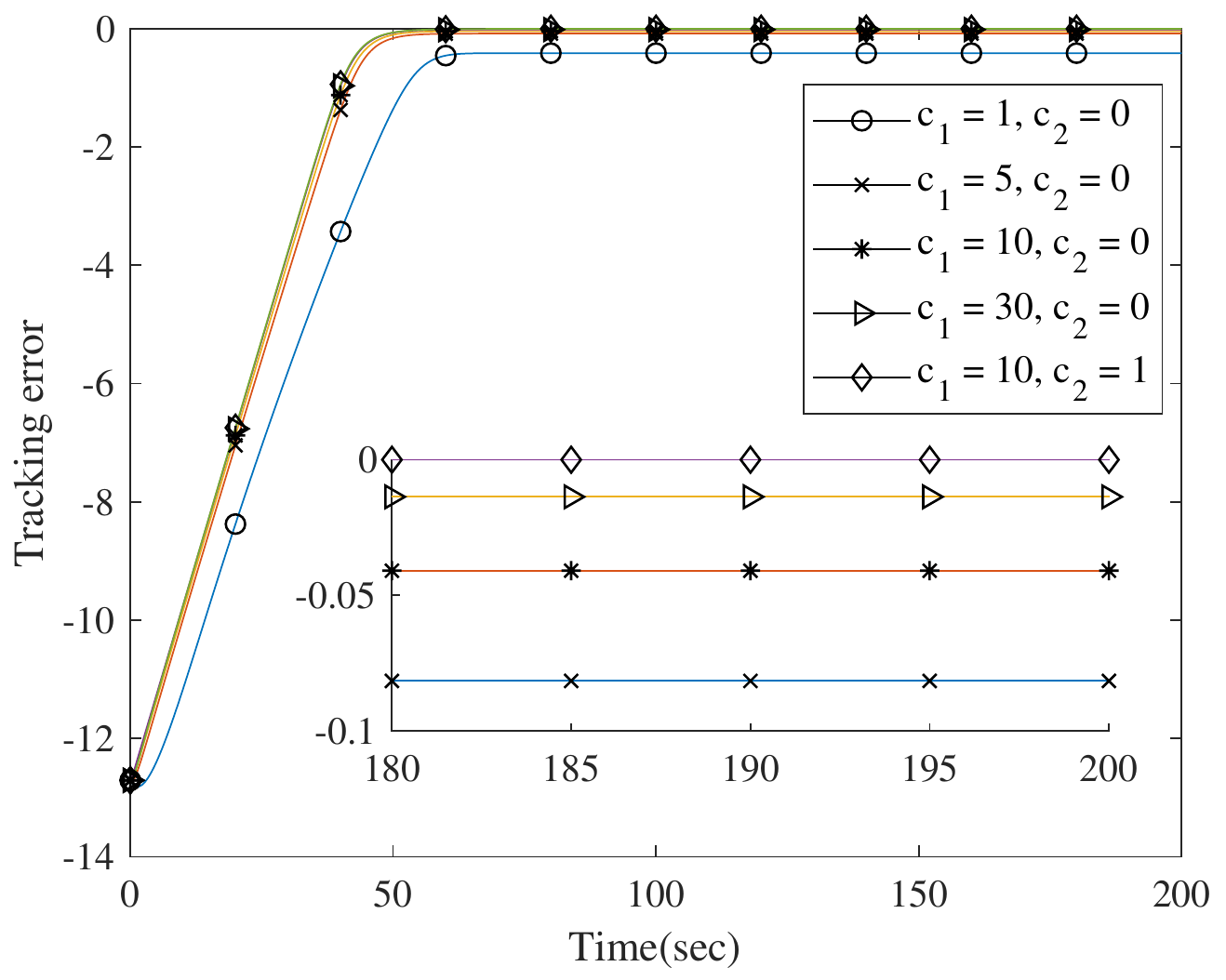}}
		\caption{Tracking errors with different control parameters.}
		\label{fig6}
	\end{figure}

\begin{figure}[t!]
		\centering{\includegraphics[width=0.8\linewidth]{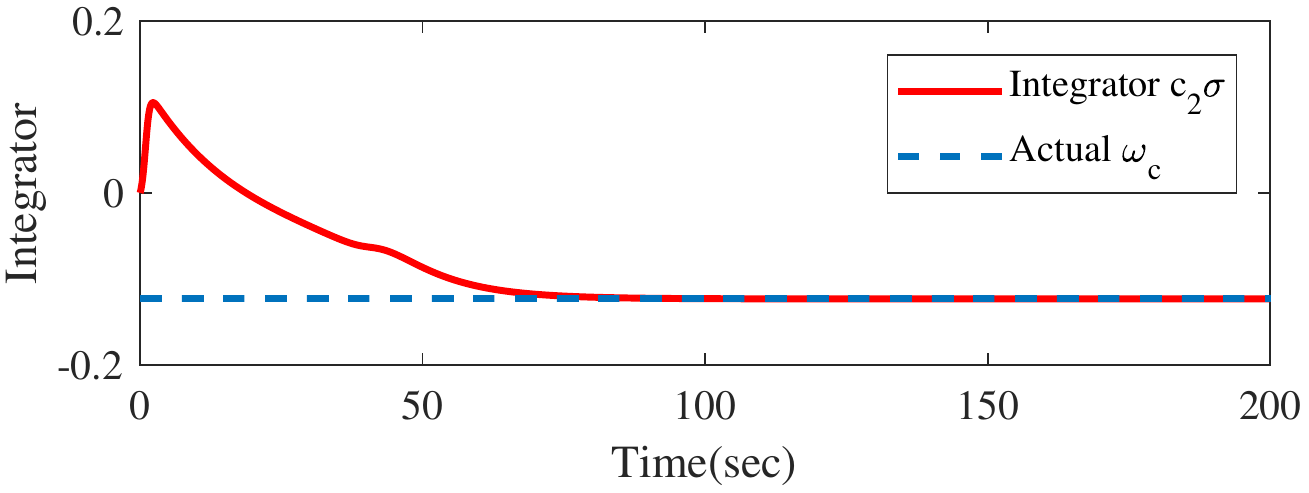}}
		\caption{The integrator $c_2 \sigma(t)$ and $\omega_c = -v/r_d$.}
		\label{fig5}
	\end{figure}

	Consider the Dubins vehicle in \eqref{eq1}, and let $\bm \beta(t) = [\bm p'(t),\theta(t)]'$ denote its state. The linear speed  is set as $v = 0.5$ \si{m/s} and the circular field of \eqref{circufield} is
	\begin{align}\label{field_circular}
		F(\bm p) = 30 \exp\left(-0.1 \sqrt{( x-5)^2 + (y-5)^2}\right).
	\end{align}
	The control parameters of the PI-like controller \eqref{eq_Dubins} is given in Table \ref{tab3}. Fig.~\ref{fig4} illustrates the field distribution of \eqref{field_circular} and the trajectories under different initial states: $\bm \beta(t_0)=[15,-5,0.6\pi]'$, $[15,15,\pi]'$, $[-5, 15, \pi/2]'$, $[-5,-5,0]'$, $[6,5,0]'$, $[5,6,\pi/2]'$, $[4,5,\pi]'$, and $[5,4,-\pi/2]'$.  Fig.~\ref{fig6} depicts the tracking errors and confirms that increasing $c_1$ can reduce the tracking error and only the PI-like controller with $c_2=1$ exactly achieves the objective in \eqref{eqobj}. Fig.~\ref{fig5} validates that the integrator $c_2 \sigma(t)$ converges to $\omega_c=-v/r_d$.
	\begin{table}[!t]
		\caption{Parameters of the PI-like controller \eqref{eq_Dubins} in Section \ref{sec_sub2}}
		\centering	
		\begin{tabular}{|c|c|c|c|c|}	
			\hline
			{Parameter}   &{$c_1$} &{$c_2$} &{$c_3$} &{$c_4$}\\
			\hline
			{Value}         & 10          & 1  & 0.3  & 1              \\       
			\hline
		\end{tabular}%
		\label{tab3}%
	\end{table}%

	\subsection{Comparison with other controllers for circumnavigation}
	\begin{figure}[t!]
		\centering{\includegraphics[width=0.8\linewidth]{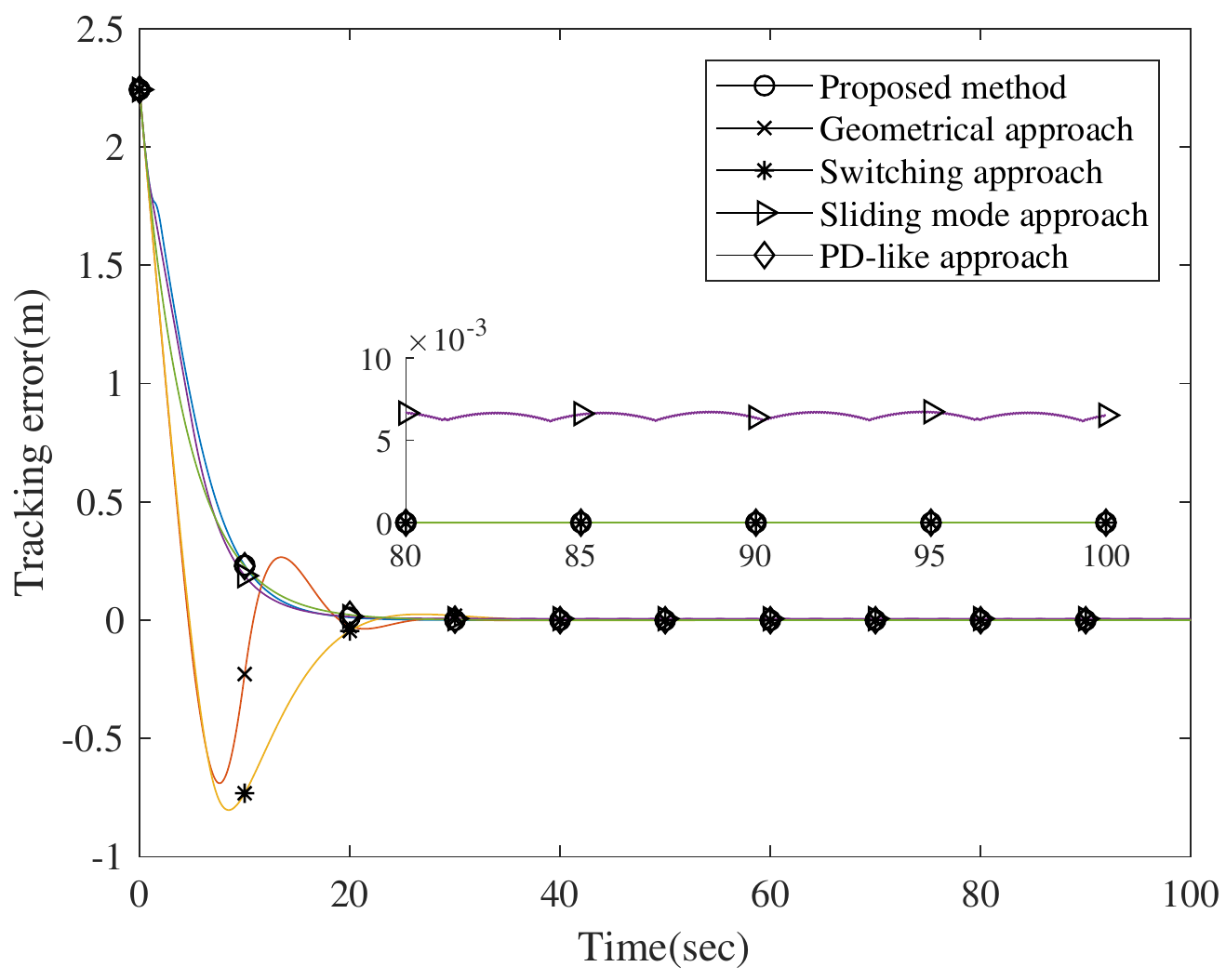}}
		\caption{Comparison with the existing methods.}
		\label{fig10}
	\end{figure}
	We compare our PI-like controller \eqref{eq_Dubins} with other methods in the context of circumnavigation including (a) the geometrical approach \citep{Cao2015UAV} with parameters $k=1$ and $r_a= 9.95$; (b) the switching approach \citep{Zhang2017Unmanned} with $k = 1.4/r_d$; (c) the sliding mode approach \cite{Matveev2011Range} with $\delta=0.83$ and $\gamma = 0.3$; and (d) the PD-like approach \citep{dong2020Circumnavigating} with $c_1 =200$ and $c_2 =30$. In Fig.~\ref{fig10}, one can observe that both the geometrical approach and the switching approach have large overshoots. The sliding mode approach cannot exactly complete the isoline tracking problem and the performance of the PI-like controller is almost of the same as the PD-like approach, which however requires to know $\omega_c=-v/r_d$ and thus cannot be applied to the isoline tracking problem of this work.

	\subsection{The isoline tracking in a smoothing field} \label{sec_sub1}	
\begin{figure}[t!]
		\centering{\includegraphics[width=0.8\linewidth]{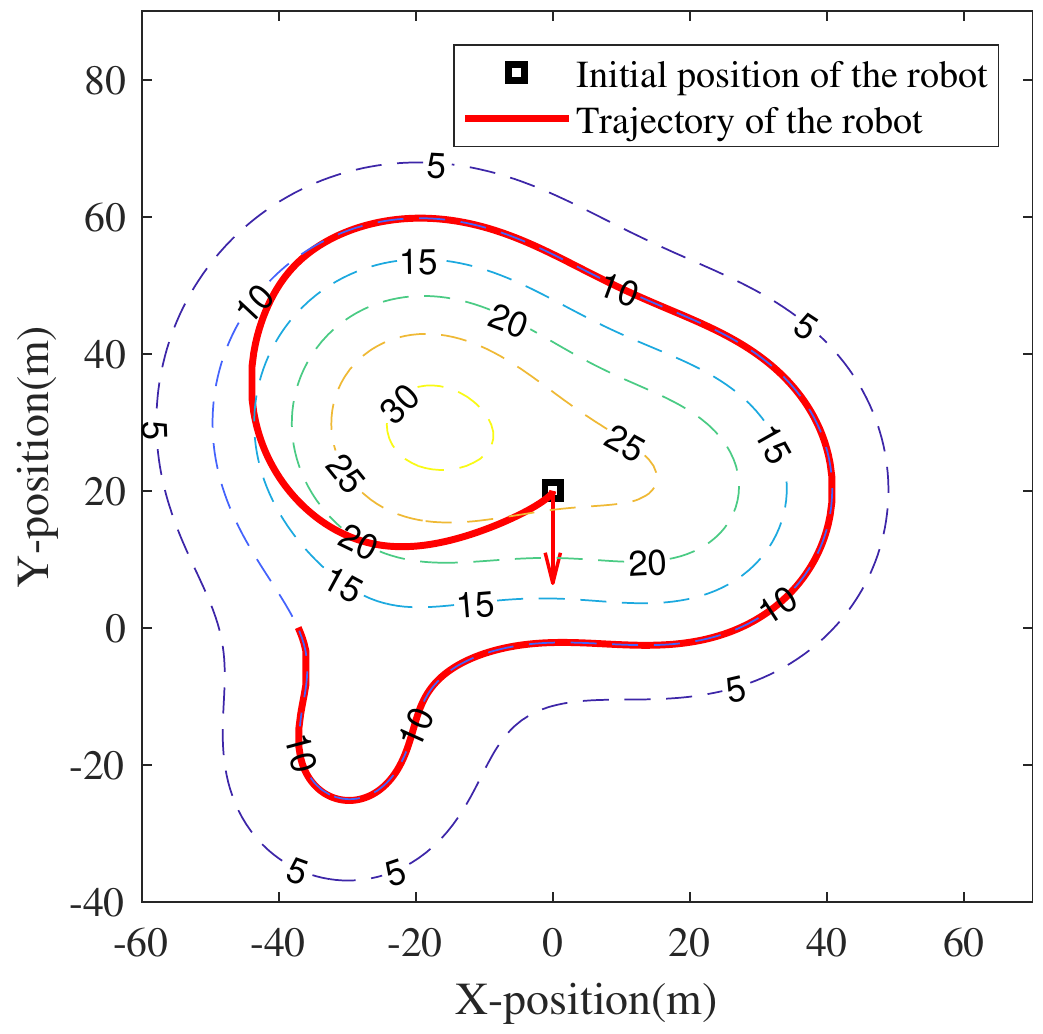}}
		\caption{Fields distribution and trajectory of the Dubins vehicle.}
		\label{fig_contour}
	\end{figure}

\begin{figure}[t!]
		\centering{\includegraphics[width=0.8\linewidth]{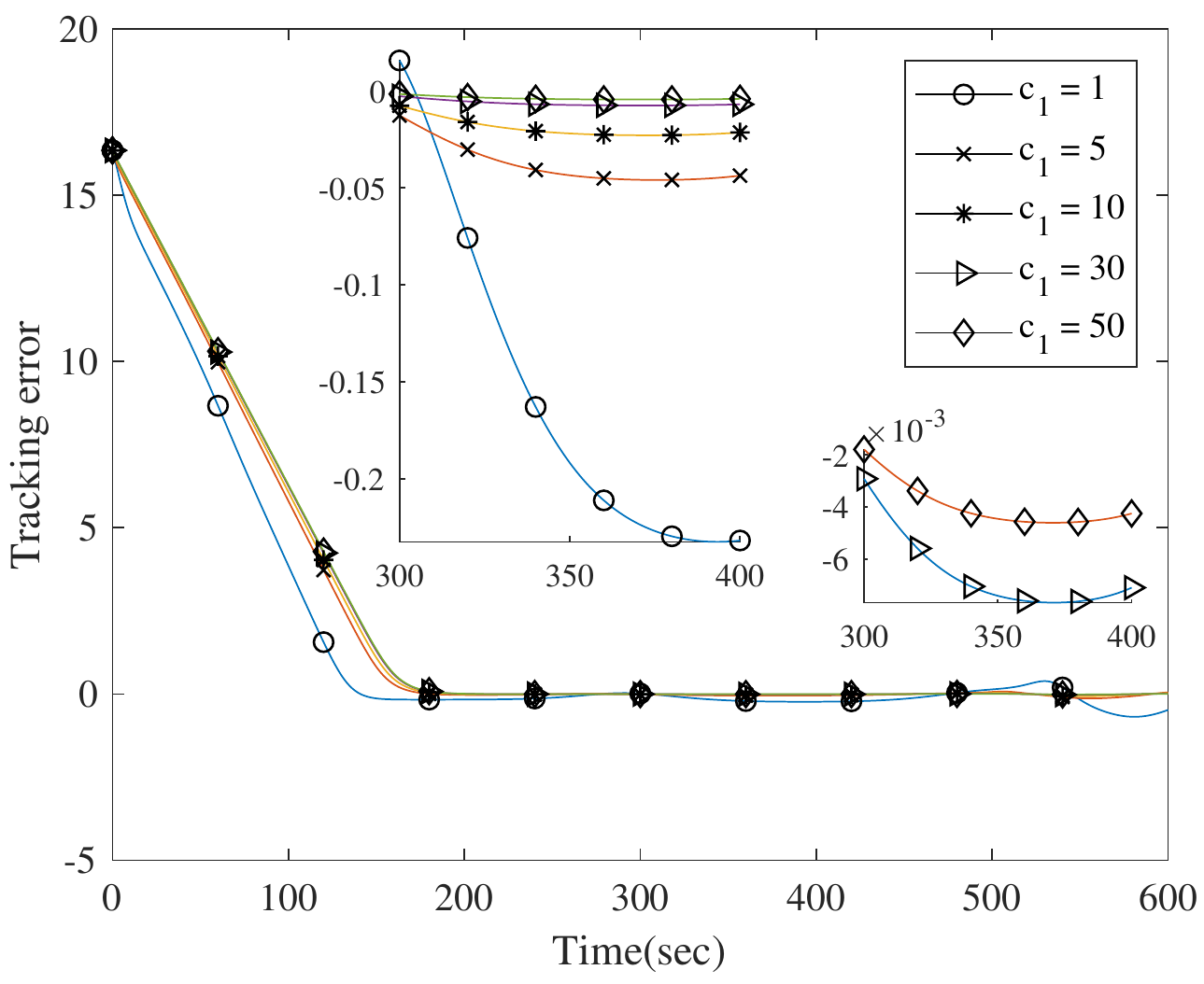}}
		\caption{Tracking errors of the Dubins vehicle with different proportional gain $c_1$.}
		\label{fig_contour_para}
	\end{figure}

\begin{figure}[t!]
		\centering{\includegraphics[width=0.8\linewidth]{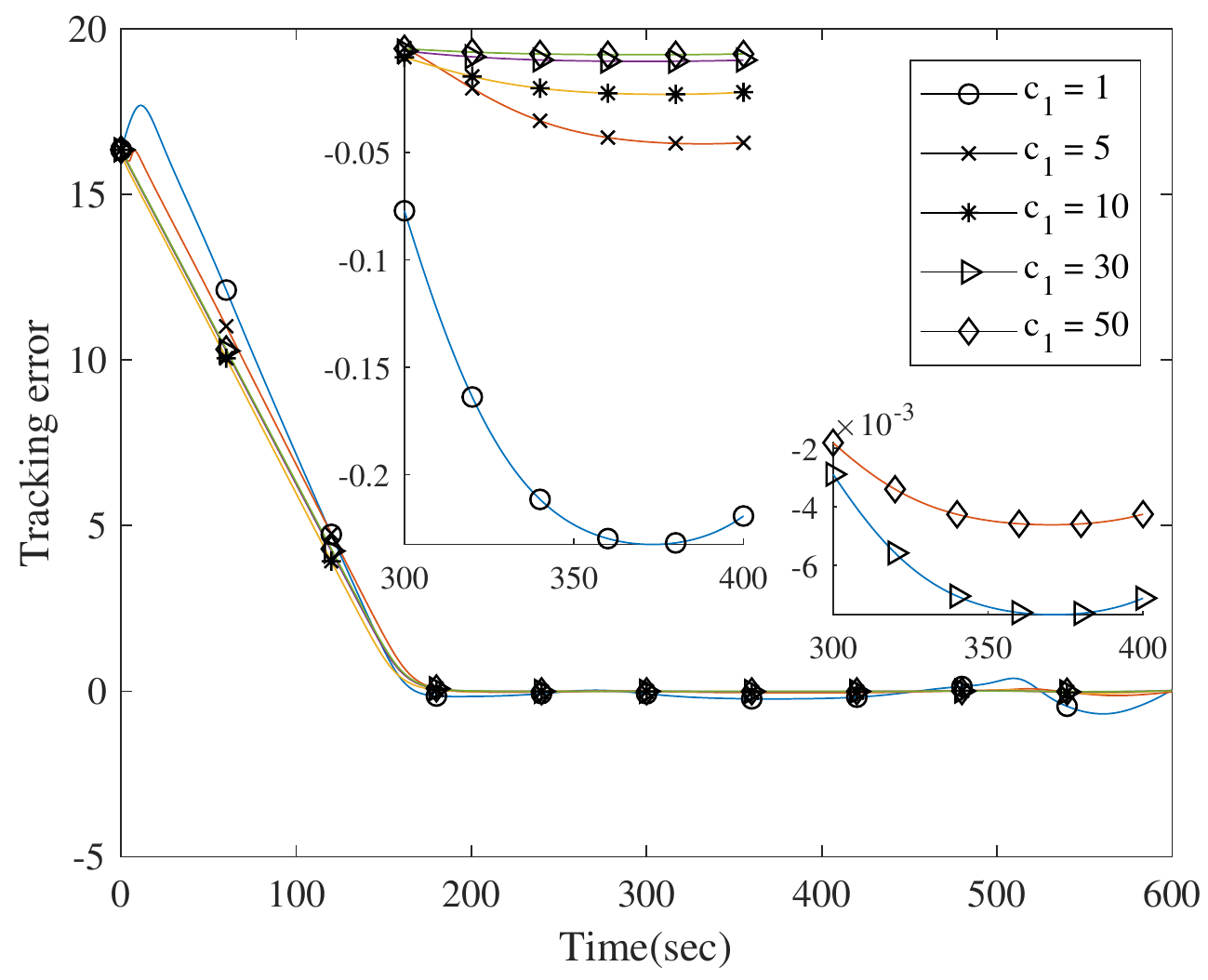}}
		\caption{Tracking errors of the single-integrator vehicle with different proportional gain $c_1$.}
		\label{fig_contour_single}
	\end{figure}

	\begin{table}[!t]
		\caption{Parameters of the PI-like controller \eqref{eq_Dubins} in Section \ref{sec_sub1}}
		\centering	
		\begin{tabular}{|c|c|c|c|c|}	
			\hline
			{Parameter}   &{$c_1$} &{$c_2$} &{$c_3$} &{$c_4$}\\
			\hline
			{Value}         & 10          & 0  & 0.1  & 1              \\       
			\hline
		\end{tabular}%
		\label{tab2}%
	\end{table}%
	 Consider the following scalar field \citep{Matveev2012Method}
		\begin{align} \label{eqexam}
		F(\bm p) =&~ 20 \exp\left( - \left((x-20)^2+(y-20)^2\right)/600 \right) +\nonumber\\
		&~30 \exp\left( - \left((x+30)^2+(y+20)^2\right)/400 \right) +\nonumber\\
		&~10 \exp\left( - \left((x+20)^2+(y-30)^2\right)/800 \right).
	\end{align} 
	The field distribution and the trajectory of the Dubins vehicle under the PI-like controller \eqref{eq_Dubins} with parameters in Table \ref{tab2} are illustrated in Fig.~\ref{fig_contour}, where $\bm \beta(t_0)=[0, 20, -\pi/2]$ and $s_d=10$. Fig.~\ref{fig_contour_para} depicts the tracking errors with  $c_1 = 1,5,10,30,50$. Clearly, we can reduce the steady-state error by increasing $c_1$ which is consistent with Proposition \ref{prop_quad}.   Fig.~\ref{fig_contour_single} validates that the single-integrator vehicle \eqref{eq_single} under the controller \eqref{controller_single} produce similar trajectory as the Dubins vehicle. 
	

\begin{figure}[t!]
		\centering{\includegraphics[width=0.8\linewidth]{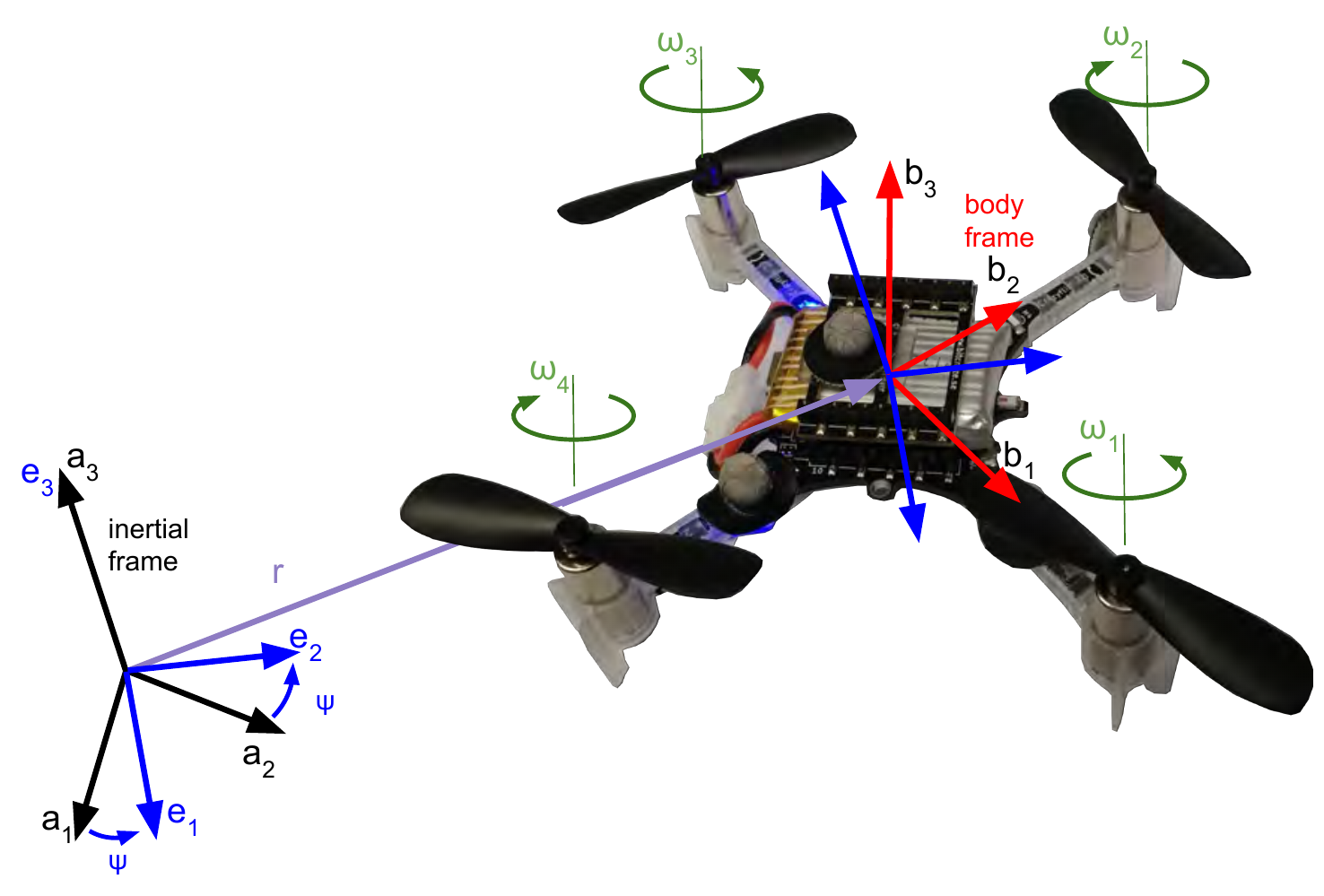}}
		\caption{CrazyFlie 2.0 quadrotor \citep{yrlu_quadrotor}.}
		\label{fig15}
	\end{figure}

\begin{figure}[t!]
		\centering{\includegraphics[width=0.8\linewidth]{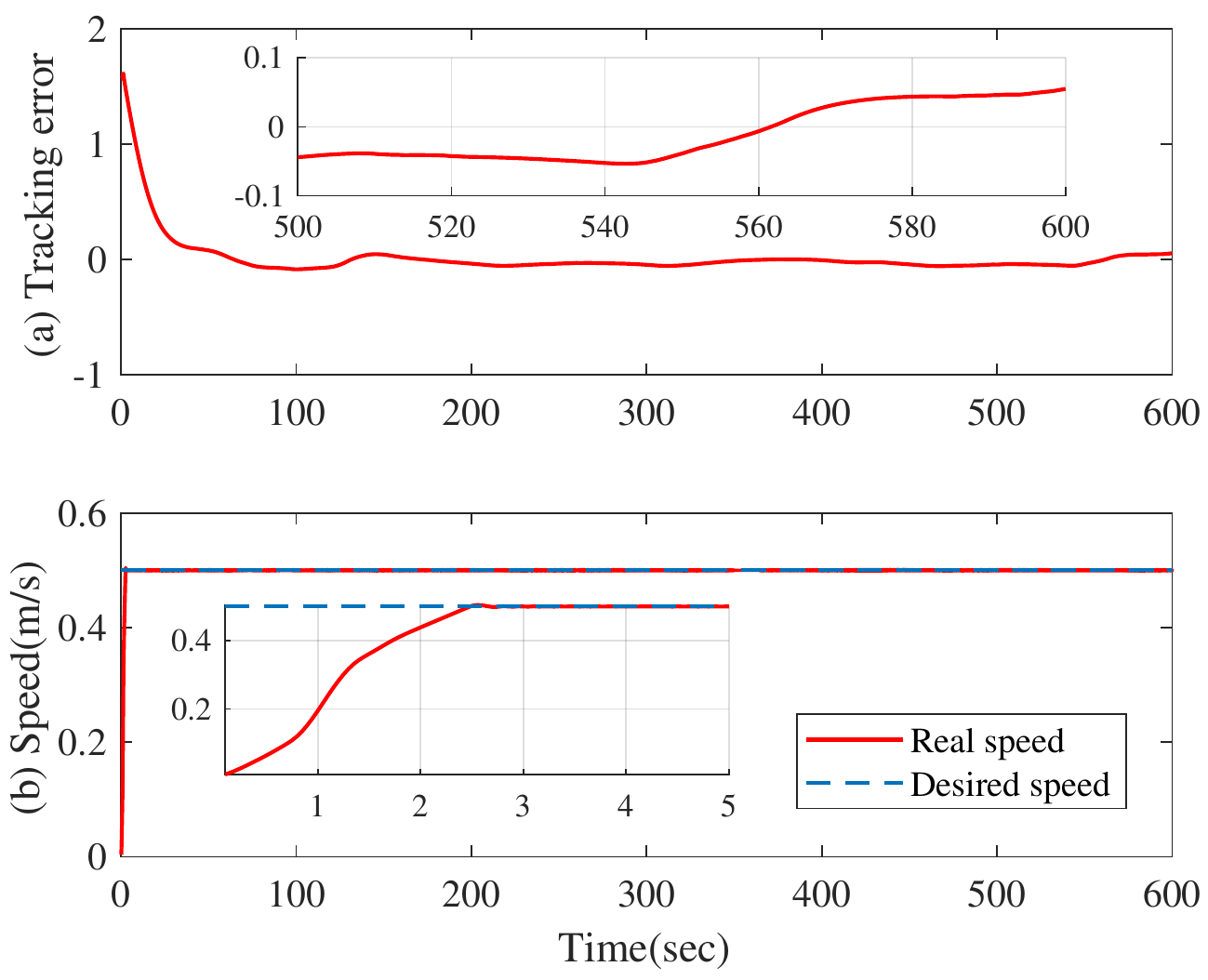}}
		\caption{Tracking error and linear speed of the quadrotor.}
		\label{fig11}
	\end{figure}

	Then, a 6-DOF quadrotor under the controller in \eqref{eq_Double} is included to complete the objective \eqref{eqobj} in the field of \eqref{eqexam}. The simulator of the quadrotor is directly obtained from \citet{yrlu_quadrotor}, which is built via CrazyFlie 2.0 platform made by Bitcraze, see Fig.~\ref{fig15} and  \citet{yrlu_quadrotor} for details. The control parameters for \eqref{eq_Double} are given in Table \ref{tab4}, and the tracking error and speed of the quadrotor are illustrated in Fig.~\ref{fig11}. Moreover, the desired isoline and speed are set as $s_d=10$ and $v=0.5$. By the partially enlarged view of Fig.~\ref{fig11}, one can observe that $v_2(t)$ converges to $v$  in a short time. Note that the altitude and attitude of the quadrotor are controlled by the original controller of \citet{yrlu_quadrotor}. 	
	\begin{table}[!t]
		\caption{Parameters of the PI-SM controller \eqref{eq_Double}}
		\centering	
		\begin{tabular}{|c|c|c|c|c|c|}	
			\hline
			{Parameter}   &{$c_1$} &{$c_2$} &{$c_3$} &{$c_4$}&{$c_5$}\\
			\hline
			{Value}         & 30          & 0.1  & 0.1  & 1 & 0.1             \\       
			\hline
		\end{tabular}%
		\label{tab4}%
	\end{table}%

	\subsection{The isoline tracking in a field of PM2.5 by a fixed-wing UAV}
\begin{figure}[t!]
		\centering{\includegraphics[width=0.8\linewidth]{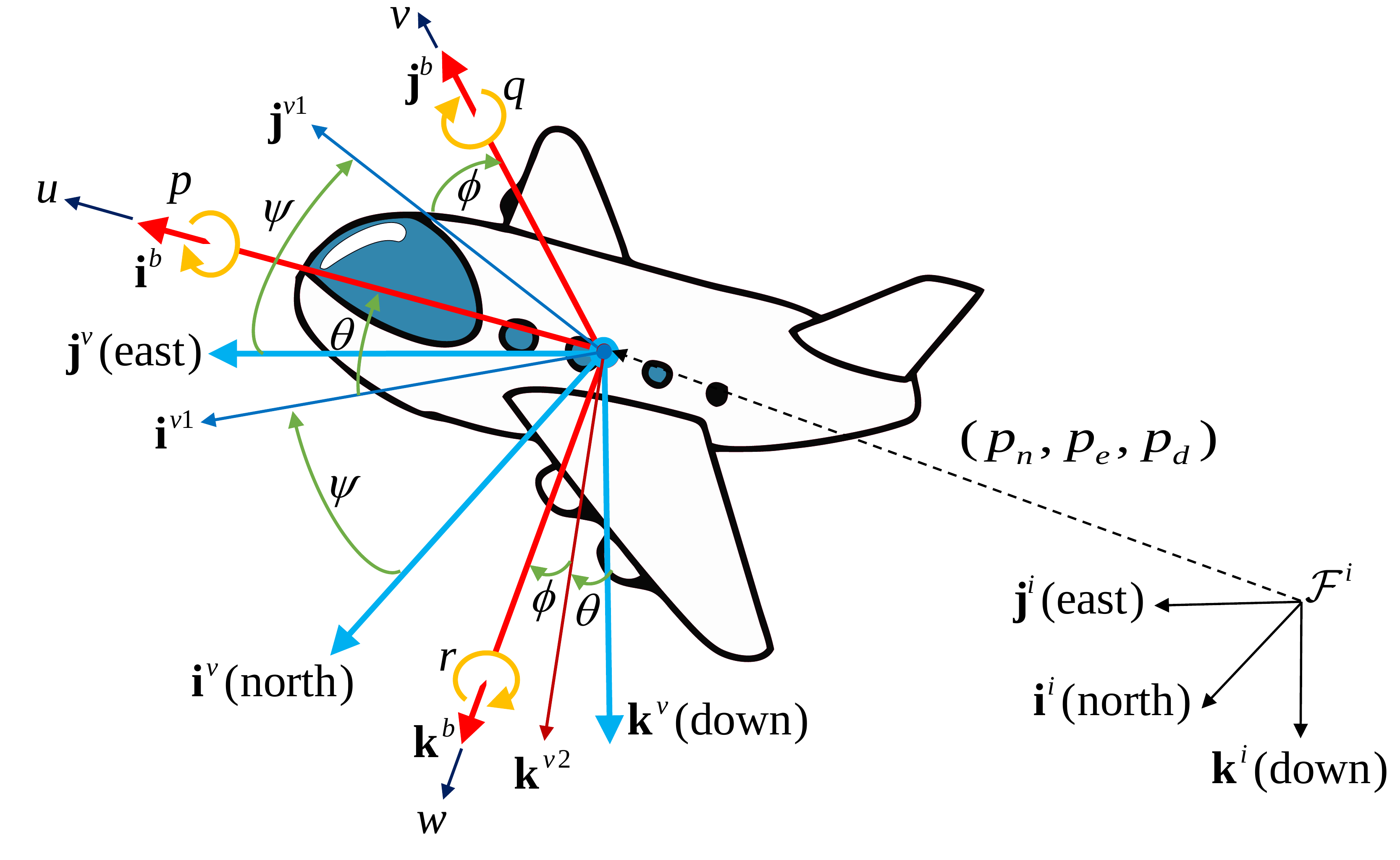}}
		\caption{Coordinates of the fixed-wing 6-DOF UAV  \citep{dong2020Circumnavigating}.}
		\label{fig7}
	\end{figure}

\begin{figure}[t!]
		\centering{\includegraphics[width=0.9\linewidth]{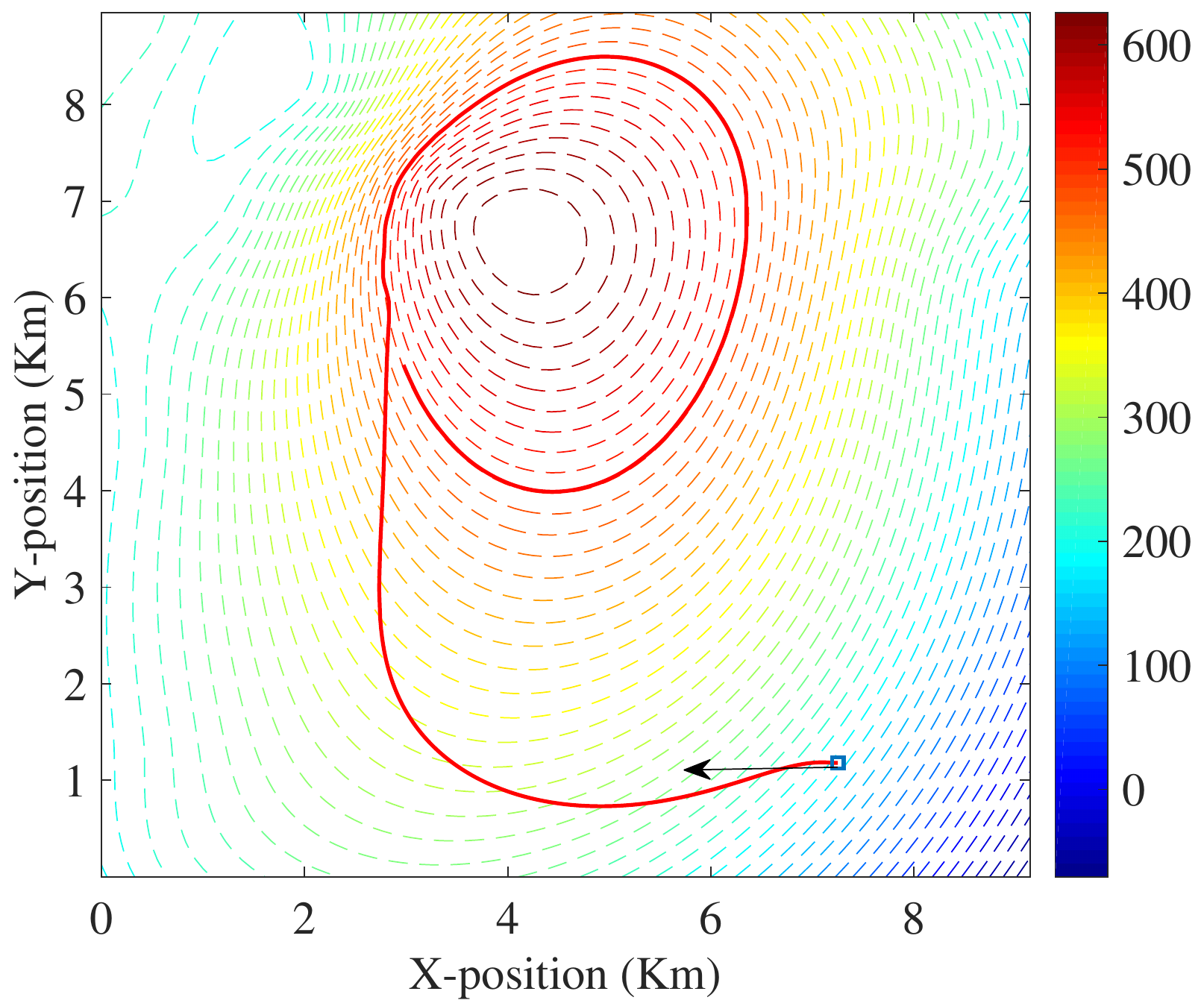}}
		\caption{Trajectory of the fixed-wing UAV in the field of PM 2.5.}
		\label{fig13}
	\end{figure}

\begin{figure}[t!]
		\centering{\includegraphics[width=0.8\linewidth]{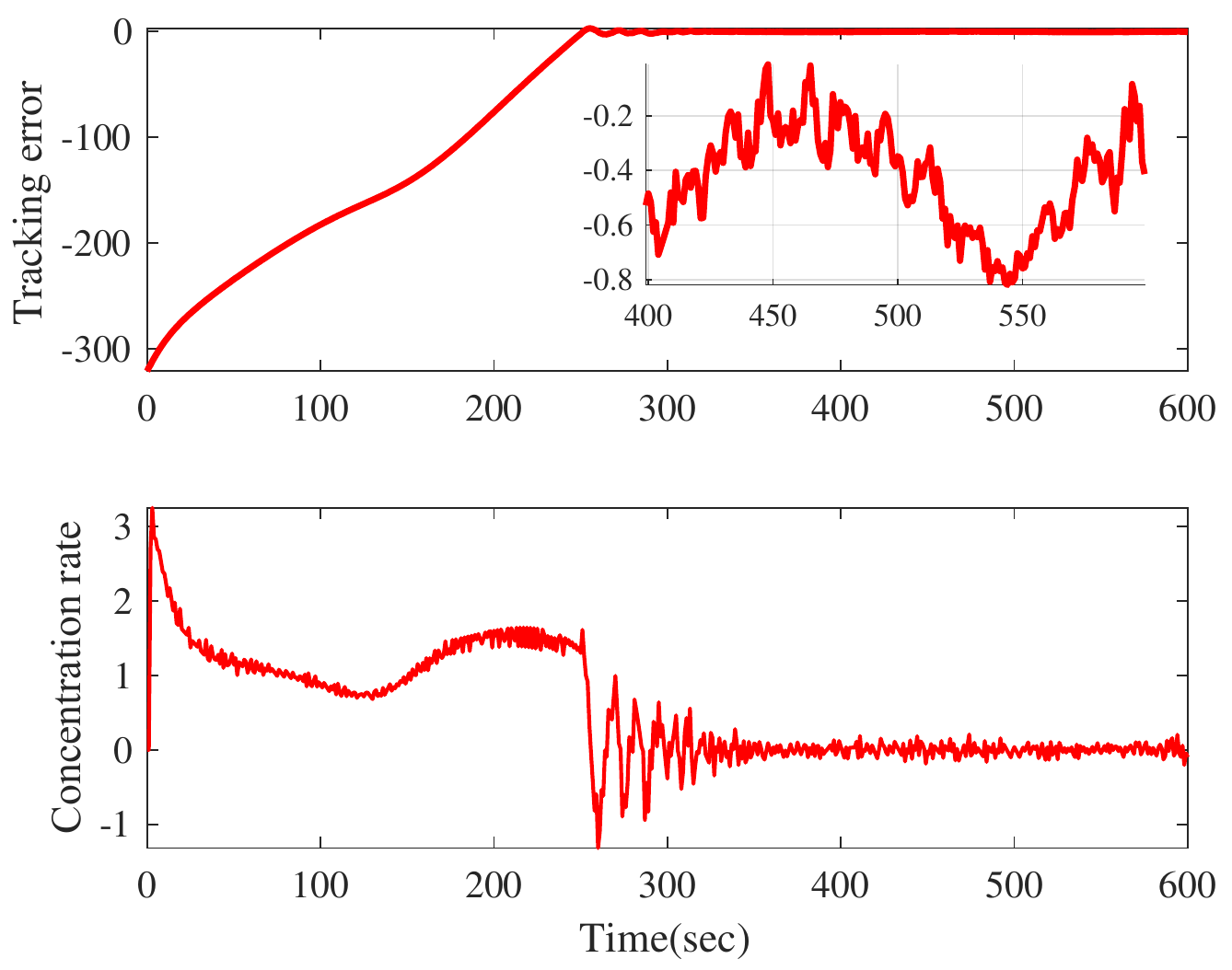}}
		\caption{Tracking error $\varepsilon(t)$ and derivative of concentration $\dot s(t)$ of the fixed-wing UAV.}
		\label{fig12}
	\end{figure}

	In this subsection, a 6-DOF fixed-wing UAV \citep{Beard2012Small} is adopted to test the effectiveness of the PI-like controller \eqref{eq_Dubins} in the field of PM2.5, see Figs.~\ref{fig1} and \ref{fig7}. 
	Due to page limitation, we omit details of the mathematical model of the UAV, which can be found in \citet{Beard2012Small}, and adopt codes from \citet{small} for the model. The sampling frequency for the PM2.5 is set as $1$ \si{Hz} and the linear speed of the UAV is maintained as $30$ \si{m/s} \citep{small}.  
Fig.~\ref{fig13} depicts the distribution of the field and the trajectory of the UAV, where the square and arrow denote its initial position and course. Fig.~\ref{fig12} illustrates the tracking error $\varepsilon(t)$ and the derivative of concentration $\dot s(t)$ versus time, which completes the isoline tracking task. 	
		
	\section{Conclusion} \label{sec6}	
	To track a desired isoline of a smoothing scalar field, we have designed a coordinate-free controller in a PI-like form for a Dubins vehicle by using concentration-based measurements in this work. A novel idea lies in the design of a sliding surface based error in the standard PI controller. Moreover, we have extended the PI-like controller to the cases of a single-integrator vehicle and a double-integrator vehicle, respectively. Finally, the simulation results have validated our theoretical finding.
	
\section*{Appendix. Proof of Proposition \ref{prop_propo}}
	To prove Proposition \ref{prop_propo}, we first show that the closed-loop system in \eqref{eqmp} asymptotically converges  if $\phi(t_0)\in [-\pi,0]$ in Lemma \ref{lemma1}. Then, we show its local exponential stability in Lemma \ref{lemma2}. Finally, we prove that there exists a finite time instant $t_1\ge t_0$ such that $\phi(t) \in [-\pi,0], ~\forall t\ge t_1$ for any initial state in Lemma \ref{lemma3}.
	
	\begin{lem} \label{lemma1}
	Under the conditions in Proposition \ref{prop_propo}, if $\phi(t_0) \in [-\pi,0]$, then 
	\begin{align}
		\lim\nolimits_{t\rightarrow \infty} |s(t)- s_e| = \lim\nolimits_{t\rightarrow \infty} |\dot s(t)|=0.
	\end{align}
	\end{lem} 
	\begin{pf}
		Similar to the proof of Proposition \ref{prop_quad}, we need to verify that $\phi(t)$ remains in the region $[-\pi,0]$, if $\phi(t_0)\in [-\pi,0]$.
		If $\phi(t) = 0$, it follows from (\ref{eqmp}) that  
		\begin{align*}  
			\dot \phi(t)=  {c_1} \left(  \alpha v + c_3 \tanh \left({\varepsilon(t)}/{c_4}\right)  \right)\ge  {c_1} ( \alpha v - c_3 ) > 0.
		\end{align*}
		Similarly, $\phi(t) = -\pi$ leads to that
		\begin{align*} 
			\dot \phi(t)  
			&\le {c_1}  (  - \alpha v + c_3 ) < 0.
		\end{align*} 
		Since $\dot\phi(t)$ is continuous in $t$, we obtain that $\phi(t) \in[-\pi,0], \forall t>t_0$. 
		
		Let $\bm y(t) =  [r(t), \phi(t)]'$ and $\bm y_e = [ r_e, -\pi/2]'$,
		where $r_e$ denotes the distance from the vehicle to the source position $\bm p_0$ if $s(t) = s_e$.
		
		Consider a Lyapunov function candidate as 
		\begin{align*}
			V(\bm y ) =& \frac{1}{v} \int_{r_e}^{y_1(t)} \left( c_1  c_3 \tanh \left(\frac{ \alpha(\tau - r_d)}{c_4}\right) -\frac{v}{\tau} \right) \text{d}\tau \\
			& + 1+\sin y_2(t).
		\end{align*}
		Taking the time derivative of $V(\bm y)$ leads to that 
		\begin{align} \label{eqvd}
			\dot V(\bm y)=&~ \left({c_1  c_3} \tanh \left(\frac{ \alpha(y_1(t) - r_d)}{c_4}\right) -\frac{v}{y_1(t)}\right) \cos y_2(t) \nonumber \\
			& + \left(\omega(t) - {v\sin y_2(t)}/{y_1(t)}\right) \cos y_2(t)  \\
			= & -v\cos y_2(t)  \left(  {c_1} \alpha \cos y_2(t) + \frac{\sin y_2(t)}{y_1(t)} + \frac{1}{y_1(t)}\right ). \nonumber
		\end{align}
		
		(a) If $y_2(t) \in [-\pi/2,0]$, then $\cos y_2(t)\ge0$ and $$\sin y_2(t)/{y_1(t)} + {1}/{y_1(t)}\ge 0,$$ which implies that $ \dot V(\bm y)\le 0.$	
		
		(b) If $y_2(t) \in [-\pi,-\pi/2)$, then $\cos y_2(t)<0$, and three cases are considered separately to check the sign of $ \dot V(\bm y)$.
		\begin{itemize}
			\item[(i)] If $y_1(t)\ge r_d$, then 
			\begin{align*}
				c_1 \alpha  \cos y_2(t) <  \cos y_2(t)/{r_d} \le  \cos y_2(t)/{y_1(t)}.
			\end{align*}
			Together with $\cos y_2(t)+ \sin y_2(t) + 1<0$, it holds that $ \dot V(\bm y)< 0$.
			\item[(ii)] If $ 1/(c_1\alpha) <y_1(t)<r_d$, then 
			\begin{align*}
				y_1(t) > \frac{1}{c_1 \alpha} > \frac{1}{c_1 \alpha} \left( \frac{1+\sin y_2(t)}{-\cos y_2(t)} \right)
			\end{align*}
			and $ \dot V(\bm y)< 0.$
			\item[(iii)] If $0<y_1(t)\le1/(c_1\alpha)$,  it follows from \eqref{eqmp} that
			\begin{align*}
				\dot y_2(t) 
				>-c_1 \alpha v \cos y_2(t)  -{c_1 \alpha v} \sin y_2(t)> c_1 \alpha v>0.
			\end{align*}
			Thus, $y_2(t)$ will monotonically increase until entering $[-\pi/2,0]$, which is Case (a). Moreover, when $y_2(t) = -\pi/2$, it holds that
			\begin{align*}
				\begin{cases}
					\dot y_2(t)<0 , & \text{if}~ y_1(t)> r_e,\\
					\dot y_2(t)=0 , & \text{if}~ y_1(t)= r_e,\\
					\dot y_2(t)>0 , & \text{if}~ y_1(t)< r_e.
				\end{cases}
			\end{align*}
			That is, the vehicle never return to  Case (iii). Finally, we have $ \dot V(\bm y)< 0.$
		\end{itemize}
		
		Let $\mathcal S=\{\bm y  | \dot V(\bm y) =0  \}$. For any $\widetilde{\bm y}_e \in \mathcal S$ and $\widetilde{\bm y}_e\neq \bm y_e$, then
		$
			\dot  y_2 |_{\bm y =\widetilde{\bm y}_e}={ c_1 c_3}\tanh\left({\varepsilon(t)}/{c_4}\right) + {v}/{y_1(t)}   \neq 0.
	        $
	        Thus, no solution can stay identically in $\mathcal S$ other than $\bm y(t) \equiv \bm y_e$. Note that $V(\bm y)$ is nonnegative, and $V(\bm y)>0$, $\forall \bm y \neq \bm y_e$. By the LaSalle's invariance theorem \citep[Corollary 4.1]{Khalil2002Nonlinear}, $\bm y_e$ is an asymptotically stable equilibrium of the closed-loop  system in \eqref{eqmodel} under the P-like controller \eqref{eqp}, i.e., 
		$
			\lim\nolimits_{t \rightarrow \infty} \bm x(t) = \widetilde{\bm x}_e,
		$
		which is implied by \eqref{eq_doubleield}.    \qed
	\end{pf}
	
	\begin{lem} \label{lemma2} Under the conditions in Proposition \ref{prop_propo},  if $\phi(t_0) \in [-\pi,~0]$, then there is a finite $t_1\ge t_0$ such that 
	\begin{align*}
		\twon{\bm x(t)-\widetilde{\bm x}_e}\le C\twon{\bm x(t_1)-\widetilde{\bm x}_e} \exp\left(-\rho (t-t_1)\right), \forall t>t_1,
	\end{align*}
	where $\rho$ and $C$ are two positive constants. 
	\end{lem}
	\begin{pf}
		Firstly, we define $\bm x(t)= [x_1(t), x_2(t)]'$ and recall the closed-loop system in (\ref{eqmp}) that
		\begin{equation} \label{eq23}
			\begin{split}
				\dot x_1(t) =&~ -\alpha v \cos x_2(t), \\
				\dot x_2(t) =&~ c_1 \left(\dot x_1(t) + {c_3}\tanh\left({(x_1(t)-s_d)}/{c_4} \right)\right) \\
				&+{\alpha v \sin x_2(t)}/{\left(\alpha r_d + s_d -x_1(t)\right)}  .
			\end{split}
		\end{equation}
	      Linearizing \eqref{eq23} around $\widetilde{\bm x}_e$ leads to that 
		\begin{align} \label{eq34}
			\dot {\bm x} (t) = \widetilde{A} (\bm x(t) -\widetilde{\bm x}_e) ~\text{and}~ \widetilde{A}=\bmatri 0 &  -\alpha v  \\a_{21}   &- {c_1 \alpha v} \ematri,
		\end{align}	
		where $$a_{21} = \frac{c_1c_3}{c_4}\left(1-\tanh^2\left(\frac{s_e-s_d}{c_4}  \right)  \right) -\frac{\alpha v}{(\alpha r_d +s_d -s_e)^2}.$$
		
		Obviously, both the eigenvalues of $\widetilde{A}$ have negative real part, i.e., $\widetilde{A}$ is Hurwitz. Let 
		\begin{align} \label{eq47}
			\mathcal D=\{ \bm x| V(\bm x)\le b \},
		\end{align}
		where $b>0$. If $b$ is sufficiently small, then $x_1(t)$ is sufficiently close to $s_e$ and $x_2(t)$ is sufficiently close to $-\pi/2$.
		
		By Lemma \ref{lemma1}, there exists a finite $t_1$ such that $\bm x(t) \in \mathcal D$ for all $t>t_1$. Then, it follows from (\ref{eq34}) that the trajectory of the system satisfies  
		\begin{align*}
			\bm x(t) -\widetilde{\bm x}_e = G\exp(\Lambda (t-t_1))G^{-1} (\bm x(t_1)-\bm x_e), \forall t> t_1,
		\end{align*}
		where $\widetilde{A} = G\Lambda G^{-1}$, $\Lambda= \diag( \lambda_1,\lambda_2 )$, and $\lambda_i$, $i=1,2$ are the eigenvalues of matrix $\widetilde{A}$. 
		Finally, it holds that
		\begin{align*}
			\twon{\bm x(t) -\widetilde{\bm x}_e } &= \twon{ G\exp(\Lambda (t-t_1))G^{-1} (\bm x(t_1)-\bm x_e)} \\
			&\le C\twon{\bm x(t_1)-\widetilde{\bm x}_e} \exp(- \rho (t-t_1)),
		\end{align*}
		where $C=\twon{G}\twon{G^{-1}}$,  $\Delta=(c_1 \alpha v)^2 -4 a_{21} \alpha v$, and 
		\begin{align*}
			\rho=
			\begin{cases}
				(c_1 \alpha v - \sqrt{\Delta})/2, &~\text{if}~ \Delta > 0,\\
				c_1 \alpha v /2, &~\text{if}~ \Delta \le 0.
			\end{cases}
		\end{align*}	        \qed
	\end{pf}

	\begin{lem} \label{lemma3}
	Under the conditions in Proposition \ref{prop_propo}, there exists a finite $t_1> t_0$ such that $\phi(t_1) \in [-\pi,0]$ for any initial state $\phi(t_0) \in (0,\pi)$.
	\end{lem} 
	\begin{pf}
		To prove Lemma \ref{lemma3}, four cases in Fig. \ref{fig16} are considered.
		
		\begin{figure}[t!]
			\centering{\includegraphics[width=1.0\linewidth]{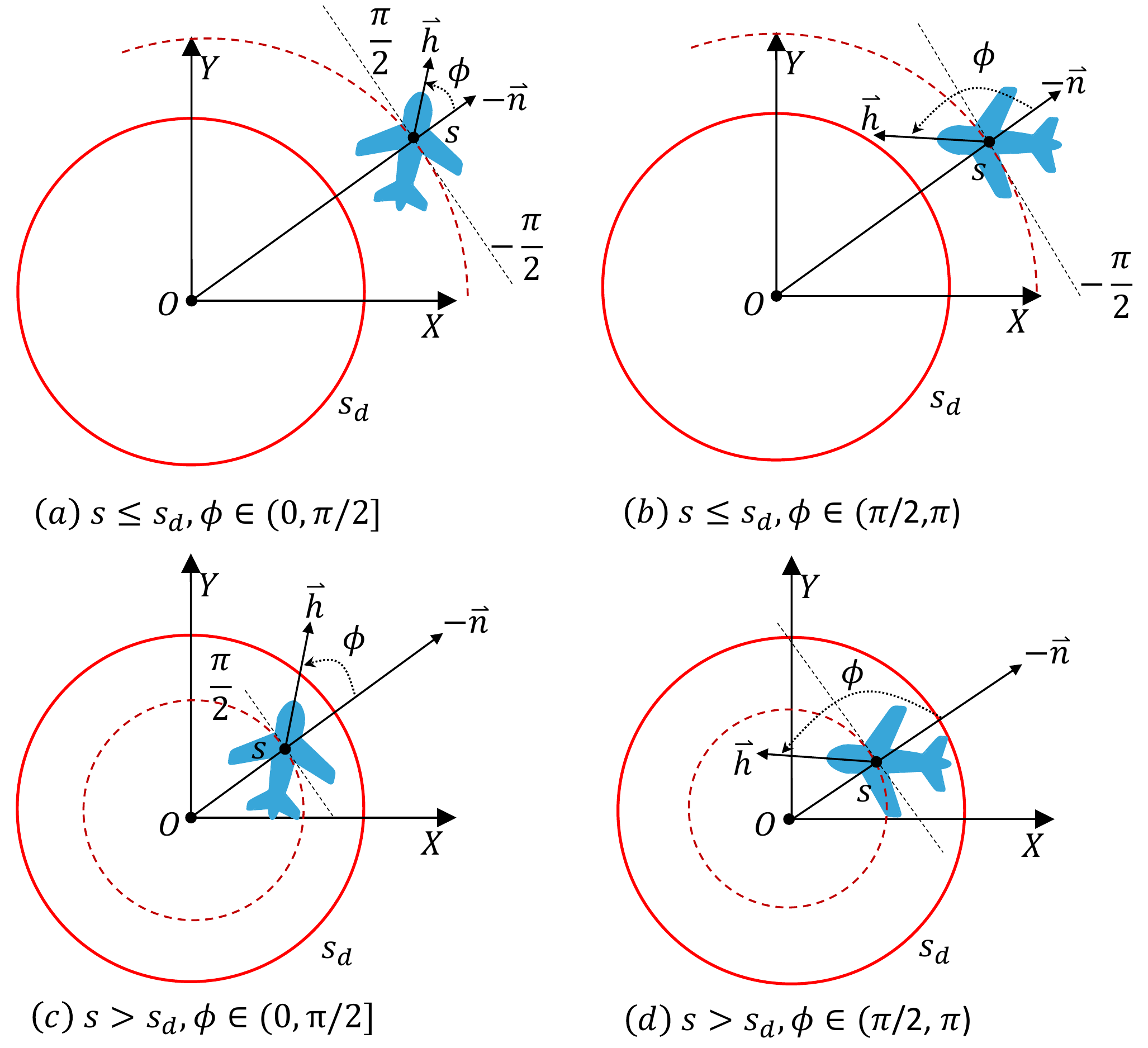}}
			\caption{Illustrations of the state of the Dubins vehicle.}
			\label{fig16}
		\end{figure}
		
		For the case in Fig. \ref{fig16}(a), i.e., $s(t_0) \in (0,s_d]$ and $\phi(t_0) \in (0,\pi/2]$,  it follows from \eqref{eqmp} that $\dot s(t_0) \le 0$ and 
		\begin{align*}
			\dot \phi(t_0) &= c_1 \left(\dot s(t_0) + c_3 \tanh \left( {\varepsilon(t_0) }/{c_4} \right) \right) - {v\sin \phi(t_0)}/{r(t_0)} \\
			& < -c_1 \alpha v \cos \phi(t_0)  <0.
		\end{align*}
		Since $\phi(t)$ is continuous in $t$, $\phi(t)$ will monotonically decrease until $\phi(t_0+\delta) \in [-\pi,0]$ where $\delta>0$ is finite. 
			
		For the case in Fig.~\ref{fig16}(b), i.e., $s(t_0) \in (0,~s_d]$ and $\phi(t_0) \in (\pi/2,~\pi)$, it follows from  (\ref{eqmp}) that 
		\begin{align*}
			\begin{cases}
				\dot \phi(t) <0, & \text{if}~ \phi(t) = \pi/2, \\
				\dot \phi(t) > 0, & \text{if}~ \phi(t) = \pi.
			\end{cases}
		\end{align*}
		Then, there are three possible results after some finite time $\delta>0$: (i) $\phi(t_0 + \delta) \ge \pi $ and $s(t_0+\delta)\le s_d$, which is equivalent to that $\phi(t_0+\delta)\ge -\pi$; (ii) $\phi(t_0+\delta) \le \pi/2$ and $s(t_0+\delta)\le s_d$, which is the case in Fig.~\ref{fig16}(a); (iii) $s(t_0+\delta)> s_d$, which corresponds to the cases in Fig.~\ref{fig16}(c) and (d). 
		
		Next, we show that $\phi(t)$ will enter the region $ [-\pi,0]$ in finite time. 		
		When $\phi(t) = \pi/2$ and $s(t)>s_d$, it follows from (\ref{eqmp}) that 
		\begin{align*}
			\dot \phi(t) 
			&	=c_1  c_3 \tanh \left( {- \alpha/{c_4} \cdot(r(t)-r_d)} \right) -{v}/{r(t)}. 
		\end{align*}
		\begin{itemize}
			\item[(a)] If there is no solution in the region $(0,r_d)$ such that $\dot \phi(t)=0$, then $\phi(t)$ is monotonic in the cases of Fig.~\ref{fig16}(c) and (d).
			\item[(b)] If there is a solution $0<r_*<r_d$ such that $\dot \phi(t)=0$, the equilibrium $\bm y_* =[r_*, \pi/2]'$ is unstable and there is no closed orbit around it.
		\end{itemize}
		
		Overall, there are two possible results after some finite $\delta>0$: (i) $r(t_0+\delta)\ge r_d $ and $\phi(t_0+\delta) \in (0,\pi/2]$, which is Fig.~\ref{fig16}(a); (ii) $\phi(t_0+\delta) \in [-\pi,0]$. Thus, we conclude that there exists a finite time instant $t_1>t_0$ such that $\phi(t_1)\in [-\pi,~0]$ for any initial $\phi(t_0) \in (0,\pi)$.
		
		To elaborate (b), we linearize \eqref{eqmp} around $\bm y_*$ as 
		\begin{align*}
			\dot {\bm y}(t) = A_* (\bm y(t) -\bm y_*) ~\text{and}~	A_*= \bmatri  0 &  -v  \\a_{21}^*    &{c_1 \alpha v }     \ematri, 
		\end{align*}
		where 
		$a_{21}^* = -\frac{c_1 c_3\alpha}{c_4 }\left(1-\tanh^2\left( \frac{\alpha (r_*-r_d)}{c_4}\right) \right)  + \frac{v}{r_*^2}.$
		It is clear that  $A_*$ at least has one unstable eigenvalue. Then, we show that there is no closed orbit around $\bm y_*$ by applying Dulac's Criterion \citep[Section 7.2]{Strogatz2018Nonlinear} and selecting a continuously differentiable, real-value function $h(\bm y)= y_1(t)$. If $y_1(t)\in (0,r_d)$ and $y_2(t) \in (0,\pi)$, it holds that 
		\begin{align*}
			\frac{\partial(h(\bm y) \dot y_1)}{\partial y_1}+ \frac{\partial(h(\bm y) \dot y_2)}{\partial y_2} =  -c_1 \alpha v y_1(t) \sin y_2(t)<0.
		\end{align*}	
		Thus, there is no closed orbit in the region $y_1(t) \in (0,r_d)$ and $y_2(t) \in ( 0,\pi)$. 	  \qed	
	\end{pf}
	
	
	{\em Proof of Proposition \ref{prop_propo}:} 
	If $\phi(t_0) \in [-\pi,0]$, it follows from Lemma \ref{lemma1} that the closed-loop system \eqref{eqmp} asymptotically converges to $\widetilde{\bm x}_e= [s_e, -\pi/2]'$. Moreover, the convergence speed near $\widetilde{\bm x}_e$ is exponentially fast by Lemma \ref{lemma2}. Finally, we show that there is a finite $t_1>t_0$ such that $\phi(t_1)\in [-\pi,0]$ for any $\phi(t_0)\in (0,\pi)$, in Lemma \ref{lemma3}.
	
	\bibliographystyle{agsm} 
	\bibliography{bib/mybib} 

\end{document}